\documentclass[
reprint,
superscriptaddress,
amsmath,amssymb,
aps,
prl
]{revtex4-2}

\usepackage{graphicx}
\usepackage{dcolumn}
\usepackage{bm}
\usepackage{physics}
\usepackage{array}

\begin{document}

\title{Realization of fermionic Laughlin state on a quantum processor}

\author{Lingnan Shen}
    \affiliation{Department of Physics, University of Washington, Seattle, WA, USA}

\author{Mao Lin}
    \affiliation{Amazon Braket, Seattle, WA, USA}
\author{Cedric Yen-Yu Lin}
    \affiliation{Amazon Braket, Seattle, WA, USA}

\author{Di Xiao}
    \email{dixiao@uw.edu}
    \affiliation{Department of Material Science and Engineering, University of Washington, Seattle, WA, USA}
    \affiliation{Department of Physics, University of Washington, Seattle, WA, USA}
    \affiliation{Pacific Northwest National Laboratory, Richland, WA, USA}
\author{Ting Cao}
    \email{tingcao@uw.edu}
    \affiliation{Department of Material Science and Engineering, University of Washington, Seattle, WA, USA}
\date{\today}

\begin{abstract}

Strongly correlated topological phases of matter are central to modern condensed matter physics and quantum information technology but often challenging to probe and control in material systems. The experimental difficulty of accessing these phases has motivated the use of engineered quantum platforms for simulation and manipulation of exotic topological states. Among these, the Laughlin state stands as a cornerstone for topological matter, embodying fractionalization, anyonic excitations, and incompressibility. Although its bosonic analogs have been realized on programmable quantum simulators, a genuine fermionic Laughlin state has yet to be demonstrated on a quantum processor. Here, we realize the $\nu = 1/3$ fermionic Laughlin state on IonQ's trapped-ion quantum computer using an efficient and scalable Hamiltonian variational ansatz with 369 two-qubit gates on a 16-qubit circuit. Employing symmetry-verification error mitigation, we extract key observables that characterize the Laughlin state, including correlation hole, bulk-edge correspondence, and topological entanglement entropy, with strong agreement to exact diagonalization benchmarks. This work demonstrates an end-to-end workflow to simulate material-intrinsic topological orders and provides a starting point to explore its dynamics and excitations on digital quantum processors.

\end{abstract}

\maketitle

\section{Introduction}

Topological phases of matter, which defy the conventional Landau symmetry-breaking paradigm, forms a foundation of modern condensed matter physics \cite{wen_quantum_2017}, underpin phenomena such as the fractional quantum Hall (FQH) effect \cite{tsui_two-dimensional_1982, wen1990ground} and quantum spin liquids \cite{wen2002quantum}. 
Beyond their fundamental significance, these topological orders play a central role in fault-tolerant topological quantum computation due to their ground-state degeneracy and anyon excitations \cite{nayak_non-abelian_2008, fowler_surface_2012, Dennis2002}.
Currently, two primary approaches exist for realizing topological order: synthetic order on quantum simulators and processors, and intrinsic order in material systems. 
The past decade has witnessed significant progress in realizing the synthetic topological order \cite{satzinger_realizing_2021, andersen_non-abelian_2023, iqbal_non-abelian_2024}, demonstrating the feasibility of noisy intermediate-scale quantum (NISQ) devices \cite{preskill_quantum_2018} as a controllable experimental platform. 
These breakthroughs have primarily relied on exactly solvable model Hamiltonians with straightforward mathematical structures, such as the toric code \cite{kitaev_fault-tolerant_2003} and the $\mathcal{D}(D_4)$ quantum double model \cite{tantivasadakarn_shortest_2023}, to construct optimal shallow circuits achievable on current NISQ devices. 

While synthetic topological orders have advanced rapidly alongside the development of NISQ devices, the quest to realize material-intrinsic topological orders, such as the FQH effect, fractional Chern insulator, and quantum spin liquid, remains largely confined to solid-state devices \cite{tsui_two-dimensional_1982, cai_signatures_2023, park_observation_2023, zeng2023thermodynamic, broholm2020quantum}. These realizations are inherently challenging due to the stringent conditions required for topological phases to emerge, including careful material selection and precise control over interactions, disorder, and temperature. The scarcity of material platforms hosting intrinsic topological order has fueled great interest in exploring such exotic phases with programmable quantum simulators \cite{clark2020observation, leonard2023realization, wang2024realization, yao2013realizing, semeghini2021probing, evered_probing_2025}.
Quantum processors, in particular, offer a unique opportunity to simulate and explore a class of many-body Hamiltonians that host material-intrinsic and topologically ordered phases, enabling access to regimes beyond current experimental reach. 
However, a major obstacle remains: a general framework that simultaneously respects the topological order and the entanglement structure—whether governed by an area or volume law—remains elusive. Unlike synthetic models, where interactions can be designed to achieve exact solvability, intrinsic topological phases in materials arise from strong electron-electron interactions that lack simple mappings to shallow quantum circuits. Overcoming this challenge requires balancing circuit efficiency, physical fidelity, and computational scalability, as realizing topological order on quantum processors necessitates deep unitary circuits to capture their defining long-range entanglement, which can quickly become infeasible on NISQ devices.

In this work, we realize the fermionic $\nu = 1/3$ Laughlin state \cite{laughlin_anomalous_1983}, a paradigmatic example of topological phases of matter, on IonQ’s trapped-ion quantum computer using a new protocol based on Hamiltonian variational ansatz (HVA). By leveraging the hierarchical structure of the Laughlin parent Hamiltonian, our ansatz construction minimizes circuit depth while preserving the symmetries of the system. This symmetry-preserving construction provides scalability, reduces classical optimization complexity, and enables symmetry-verification protocol for error-mitigation, making it especially suitable for hardware implementations. We successfully prepare the fermionic Laughlin state on a 16-qubit system with 369 two-qubit gates. 
We verify the successful preparation by directly measuring, on the quantum processor, the characteristic microscopic and topological diagnostics of the Laughlin state, including bulk-edge density structure, correlation holes, and topological entanglement entropy, which show strong agreement with exact-diagonalization (ED) benchmarks. 
We deem the preparation successful only when these independent diagnostics are simultaneously satisfied, providing mutually reinforcing evidence for the target topological phase.
This suite of FQH-specific, observable-centric criteria provides a problem-tailored benchmark for future simulations in regimes without classical ground truth, enabling digital quantum processors to make genuinely predictive statements about competing strongly correlated phases and their emergent properties. This work thus represents the first realization of a fermionic $\nu=1/3$ Laughlin state on a digital quantum processor using an end-to-end workflow. Our synergistic integration approach of Hamiltonian design, ansatz construction, and error mitigation strategy establishes a concrete workflow for digital simulations of strongly correlated topological matter and opens a route to harnessing topological orders for both fundamental physics research and quantum-information applications.

\section{The Model}
\begin{figure}
    \centering
    \includegraphics[width=0.95\linewidth]{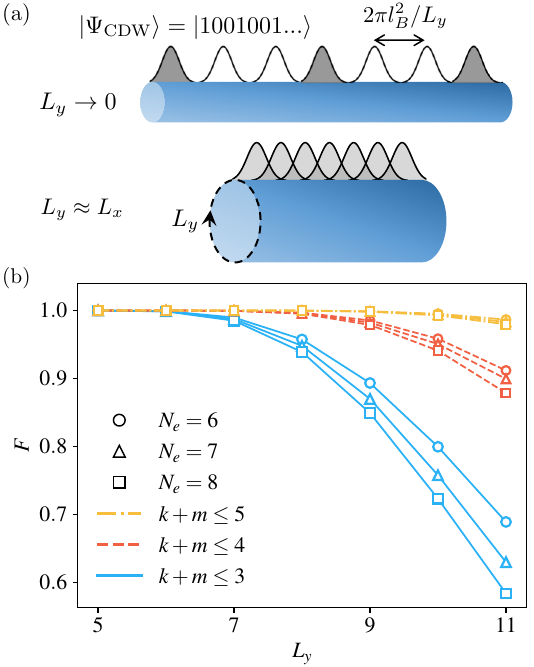}
    \caption{\textbf{Cylinder geometry and interaction truncation effect on Laughlin state. (a)} Schematic of cylinder geometries in Tao-Thouless (thin-cylinder) limit $L_y \to 0$ and the isotropic geometry limit $L_x \approx L_y$ corresponding to $L_y \approx 10$ in (b). The Gaussian peaks illustrate the localized orbitals of the lowest Landau level along the axial direction, with spacing $2 \pi l_B^2 /L_y$ where $l_B$ is the magnetic length. 
    Opacity of the Gaussian peaks represent local electron density.
    \textbf{(b)} Fidelity between the $\nu = 1/3$ Laughlin state and the ground state of the effective Hamiltonian for various truncation ranges of interactions $(k + m \le 3, 4,$ and $5)$ in Eq.~\eqref{eqn:LaughlinExactGSHam} for system with number of electrons $N_e = 6, 7,$ and $8$. 
    The cylinder's height $L_x$ is determined through the constraint $N_\Phi = L_xL_y/(2 \pi)$ where $N_\Phi$ is the number of flux quanta in the system and satisfies $N_\Phi = 3 N_e - 2$ (see Methods).
    Lines are guide to the eye.}
    \label{fig:wfc_overlap}
\end{figure}

We realize the topologically ordered Laughlin state on a quantum processor through constructing a HVA for its parent Hamiltonian defined by the following effective one-dimensional fermion chain model \cite{seidel_incompressible_2005, moudgalya_quantum_2020} on a cylinder geometry (see Methods)
\begin{equation}
    \label{eqn:LaughlinExactGSHam}
    H = \sum_j \sum_{k>m} V_{km} c^\dagger_{j+m}c^\dagger_{j + k} c_{j + k + m} c_{j},
\end{equation}
where $c^\dagger_j$ and $c_j$ are the fermionic creation and annihilation operators corresponding to the single-particle orbitals under the Landau gauge. Physically, the index $j$ specifies the x-coordinate of Gaussian-localized electron wave functions (Fig. \ref{fig:wfc_overlap}(a)). The interaction matrix elements $V_{km}$  implement the Haldane-Trugman-Kivelson pseudopotential \cite{haldane_fractional_1983, trugman_exact_1985}, under which the $\nu = 1/3$ Laughlin state is an exact ground state. This repulsive interaction decays at different rates for different interaction ranges $(k+m)$ as the cylinder's circumference $L_y$ increases. 

It is important to recognize that the Laughlin state’s defining behaviors, such as incompressible quantum liquid correlations and long-range entanglement, are not universally captured by the ground state of Eq.~\eqref{eqn:LaughlinExactGSHam} for arbitrary $L_y$. Its characteristics are hosted by the ground state of Eq.~\eqref{eqn:LaughlinExactGSHam} only near the isotropic geometry limit when the cylinder's circumference ($L_y$) matches its height ($L_x$)~\cite{seidel_incompressible_2005}. Strong deviations from it, such as the Tao-Thouless (TT) limit ($L_y \to 0$), where the ground state becomes a charge-density-wave (CDW) state $\ket{\Psi_{\text{CDW}}} = \ket{100100100...}$ (Fig. \ref{fig:wfc_overlap}(a)), and the squeezed cylinder limit ($L_y \to \infty$), where the system is collapsed into a one-dimensional Luttinger liquid, lead to unfaithful description of Laughlin state’s physical behavior.

Due to the two-body interactions in Eq.~\eqref{eqn:LaughlinExactGSHam}, the full Hamiltonian $H$ contains $\mathcal{O}(N^3)$ terms for $N$ orbitals, making variational ansatz based on the full Hamiltonian impractical for large system sizes. To address this, we develop an efficient and scalable protocol that constructs a HVA with an effective Hamiltonian $H_{\text{eff}}$ which retains only the dominant terms for correlated topological electronic systems (see Methods). 

In this protocol, the terms in $H_{\text{eff}}$ are selected and validated following two criteria: (i) quantitative fidelity of wavefunction, and (ii) qualitative preservation of topology, entanglement, and symmetry. The first criteria is universal for quantum simulations of molecules and solids. The terms in $H_{\text{eff}}$ may be identified heuristically by their large $|V_{km}|$, which determine the term’s energy scale. Their validity can be further verified via ED within computationally viable regimes, by comparing the wavefunction overlap and low-energy spectra of $H_{\text{eff}}$ and $H$. The second criteria is specific for the topologically ordered states. Qualitatively, we ensure the target state retains its defining properties—such as symmetry and topological order by verifying that $H_{\text{eff}}$ belongs to the same topological class as $H$, using topological invariants, entanglement entropy, or symmetry classifications.

Since FQH states are governed by short-range correlations, we expand Eq.~\eqref{eqn:LaughlinExactGSHam} by interaction range $(k + m)$ and evaluate the fidelity $\mathcal{F}$, defined as the wavefunction overlap between the Laughlin state and the ground state of $H_{\text{eff}}$ consisting of truncated interactions 
as a comparative diagnostic across truncation ranges, rather than as an absolute threshold. This quantifies how well $H_{\text{eff}}$ captures the Laughlin state's key features.
Fig. \ref{fig:wfc_overlap}(b) shows that from the TT limit to $L_y < 7$, all truncations regardless of the interaction range yield high fidelity. But as we approach the isotropic geometry regime $L_y \approx 10$, the Laughlin state's strong correlation and long-range entanglement kicks in. As a result, $\mathcal{F}$ drops at significantly different rate depending on the truncations range. With only the lowest-order scattering $(k + m \le 3)$, $\mathcal{F}$ drops to $0.8$ at $L_y = 10$ for system with number of electrons $N_e = 6$, whereas including longer-range interactions $(k + m \le 4,5)$ increases $\mathcal{F}$ to 0.95 and essentially 1.0, respectively.

Following the second criterion, we study how the interaction truncation range affects topology and entanglement. With only the lowest-order scattering $(k+m \le 3)$ included, the action of the effective Hamiltonian $H_{\text{TT}}$ on the CDW state $|\Psi_{\text{CDW}} \rangle$ forms a Krylov subspace $\mathcal{K}(H_{\text{TT}}, |\Psi_{\text{CDW}} \rangle)$. As an example of Hilbert space fragmentation \cite{Moudgalya_2022}, this can be used to map FQH model, such as the Laughlin state's parent Hamiltonian, under TT limit onto exactly solvable spin models \cite{rahmani_creating_2020, voinea_deformed_2024}. This Krylov subspace $\mathcal{K}$ is significantly \textit{smaller} than the full Hilbert space of a generic Laughlin state. 
As a result, the second R\'enyi entanglement entropy $S_A^{(2)}=-\ln \text{Tr}\rho_A^2$ of the $H_{\text{TT}}$ ground state, computed for a subsystem $A$ of the cylinder, rapidly saturates to a finite value as the subsystem boundary $L_y$ increases toward the isotropic limit. This behavior signals a breakdown of area law scaling and the loss of the Laughlin state's correlation structure.
In contrast, extending the truncation range to $(k + m \le 4)$ or higher restores the linear scaling of $S_A^{(2)}$ with $L_y$, recovering the expected area law behavior of a topological quantum liquid (See Supplementary Information).

Based on the quantitative criteria of fidelity and qualitative criteria of topology and entanglement, we choose $k+m \le 4$ as the truncation range of interactions in $H_{\text{eff}}$. While incorporating longer-range interactions ($k + m \ge 5$) can marginally improve fidelity, it does not qualitatively affect the topology or entanglement properties of the ground state. On the other hand, it significantly increases the complexity of the HVA circuit, pushing it beyond the capabilities of current NISQ devices. Thus we conclude the minimal $H_{\text{eff}}$ for constructing the HVA for the $\nu = 1/3$ Laughlin state includes the following interaction terms
\begin{equation}
    \label{eqn:isoHamiltonian}
    \begin{split}
        H_{\text{eff}}& = \sum_j [V_{10} \hat{n}_j\hat{n}_{j+1} + V_{20} \hat{n}_j\hat{n}_{j+2}  + V_{30} \hat{n}_j\hat{n}_{j+3} \\ 
        & + (V_{21} c^\dagger_{j+1}c^\dagger_{j + 2} c_{j + 3} c_{j} + V_{31} c^\dagger_{j+1}c^\dagger_{j + 3} c_{j + 4} c_{j} + \text{H.c.})],
    \end{split}
\end{equation}
where $\hat{n}_j =  c_j^\dagger c_j$ is the density operator.
We note that at the interaction range $k+m = 4$, we retain only the off-diagonal scattering term $V_{31}$ in $H_{\text{eff}}$, which plays a crucial role in shaping the wavefunction structure and avoiding Hilbert space fragmentation.
In contrast, $V_{40}$, despite falling within the same interaction range, is a diagonal electrostatic term that primarily results in energy shifts without significantly influencing the wavefunction. To further reduce circuit depth, we exclude $V_{40}$ from $H_{\text{eff}}$ (see Supplementary Information).

\section{Quantum Circuit for State Preparation}

With $H_\text{eff}$ identified based on our selection criteria, we construct the corresponding state preparation circuit in HVA fashion to simulate the Laughlin state on a quantum processor
with the expected HVA repetition $p$ scaling linearly with the system size; the number of variational parameters per repetition being constant, so the total parameter count scales as $\mathcal{O}(p)$.

We interpret the HVA as a digitized adiabatic protocol generated by a local effective Hamiltonian \cite{wecker_progress_2015}. Lieb–Robinson bounds on the spread of correlations under local dynamics imply an effective light cone with finite velocity \cite{bravyi_lieb-robinson_2006}. We therefore expect that, for our Laughlin state HVA, the number of repetitions $p$ must grow at least linearly with the system size in order to faithfully reproduce the long-range entanglement structure of the topological phase. 
As we show below, our ansatz also achieves a linear scaling of total number of variational parameters by generalizing parameters in a HVA layer across the lattice. 
This avoids the quadratic or worse parameter growth that would result from assigning independent parameters to every microscopic term and aligns with previous work showing that constrained HVA remains expressive while improving trainability \cite{wecker_progress_2015, park_hamiltonian_2024, mele_avoiding_2022}.

\begin{figure}
    \centering
    \includegraphics[width=0.93\linewidth]{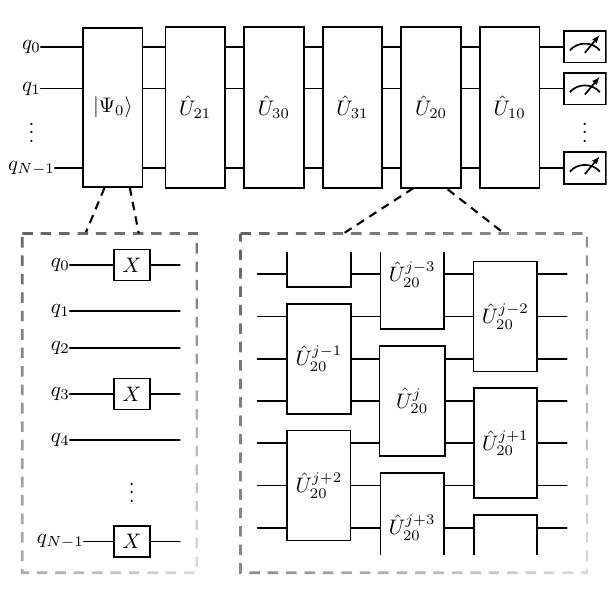}
    \caption{\textbf{Schematic $N$-qubit Hamiltonian variational ansatz circuit for preparing the $\nu = 1/3$ Laughlin state.} The initial state is taken as the charge-density wave state $\ket{\Psi_0} = \ket{100100....1001}$, where we use the Jordan-Wigner transformation in this work \cite{jordan_uber_1928}. Commuting operators in $\hat{U}_{km}$ are executed in parallel. We show the structure of $\hat{U}_{20}$ layer as an example (see Supplementary Information for a full state preparation circuit at $N_e = 6$).}
    \label{fig:circuit_schematic}
\end{figure}

The state preparation circuit $\ket{\psi(\{\beta_j\})}_{\text{eff}}$, shown in Fig. \ref{fig:circuit_schematic}, is given by the following unitaries
\begin{equation}
    \label{eqn:unitary_km}
    \hat{U}_{km} = \prod_{j} \text{exp}[-i \beta_{km} (c^\dagger_{j+m}c^\dagger_{j + k} c_{j + k + m} c_{j} + \text{H.c.})],
\end{equation}
where $\beta_{km}$ are variational parameters. The sum of indices are implicitly bound by the system size. 
The construction and optimization of $\ket{\psi(\{\beta_j\})}_{\text{eff}}$ is guided by two fundamental principles. Firstly, we generalize the variational parameters $\beta_{km}$ throughout the lattice, due to the similarity in mathematical structures at different $j$ [Eq.~\eqref{eqn:LaughlinExactGSHam}]. 
In practice, this means that all gates within the same unitary $\hat U_{km}$ share the same parameter $\beta_{km}$, yielding a constrained HVA \cite{wecker_progress_2015, park_hamiltonian_2024} with one variational parameter per physical generator $\hat U_{km}$ rather than one per microscopic term. As a result, each HVA repetition uses five independent parameters, independent of the system size.
This dimensionality reduction of parameter space not only simplifies the variational optimization but also
ensures the total number of parameters grows only through the number of HVA repetitions $p$, i.e., $N_\mathrm{param} \sim \mathcal{O}(p) \propto N_e$ (see Supplementary Information for explicit circuit gate and depth).
Secondly, the squeezing rule in FQH \cite{bernevig_model_2008} requires $\hat{U}_{21}$ as the first layer of the circuit which only contains terms with $j = 3n, n\in \mathbb{Z}$.

\begin{figure}
    \centering
    \includegraphics[width=1.0\linewidth]{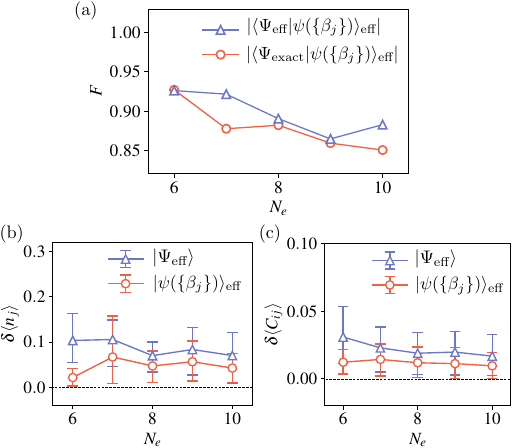}
    \caption{\textbf{Finite-depth scaling of fidelity and intensive quantities for the optimized protocol in the isotropic geometry regime. (a)} 
    Fidelity between the state preparation circuit and ground state obtained by ED for system with number of particle $N_e = 6$ to 10. (Blue triangle) Fidelity between $\ket{\psi(\{\beta_j\})}_{\text{eff}}$ and $\ket{\Psi_\text{eff}}$, ground state of $H_\text{eff}$. (Red circle) Fidelity between $\ket{\psi(\{\beta_j\})}_{\text{eff}}$ and the Laughlin state $\ket{\Psi_\text{exact}}$.
    \textbf{(b)} Average deviation of local density $\delta \langle n_j \rangle$. \textbf{(c)} Average deviation of two-point correlation function $\delta \langle C_{ij} \rangle$. In \textbf{(b)} and \textbf{(c)}, deviation of the quantity $\langle x \rangle$ is defined as $\delta \langle x \rangle = |\langle x \rangle' - \langle x \rangle_\text{exact}|$, where $\langle x \rangle_\text{exact}$ is the exact value for the Laughlin state and $\langle x \rangle'$ corresponds to $\ket{\Psi}_{\text{eff}}$ or $\ket{\psi(\{\beta_j\})}_{\text{eff}}$. All error bars indicate the 16th and 84th percentiles. Lines are guide to the eye.}
    \label{fig:fidelity_scaling}
\end{figure}

Using classical simulator (noiseless), we optimize $\beta_{km}$ for $\nu=1/3$ Laughlin state in the isotropic geometry regime (see Methods), and demonstrated that the optimized parameters obtained with $N_e = 6$ can be transferred to larger systems as warm starts, assuming a fixed HVA repetition $p$. 
The optimized parameters $\beta_{km}$ achieves $\mathcal{F}=0.93$ compared with the Laughlin state, the exact ground state of the full Hamiltonian \eqref{eqn:LaughlinExactGSHam} obtained by ED at $N_e = 6$.
Since the fidelity between the ground state of $H_{\text{eff}}$ and the Laughlin state decays naturally with system size $N_e$ (Fig. \ref{fig:wfc_overlap}), we expect the fidelity between $\ket{\psi(\{\beta_j\})}_{\text{eff}}$ and the Laughlin state to follow the same trend when we transfer the optimized parameters to larger systems. Fig. \ref{fig:fidelity_scaling}(a) shows the fidelity scales as expected for larger systems up to $N_e = 10$. Optimizing $\ket{\psi(\{\beta_j\})}_{\text{eff}}$ with larger system size did not achieve higher fidelity (see Supplementary Information), further supporting parameter transferability and our construction's resilience to barren plateau \cite{mele_avoiding_2022}.
This smooth transferability suggests that parameters optimized on smaller systems provide high-quality warm starts for larger systems, reducing classical optimization costs and mitigating trainability issues when one subsequently increases the HVA repetition $p$ with system size.

Notably, the average deviation of intensive quantities, such as the local density and two-point correlation between $\ket{\psi(\{\beta_j\})}_{\text{eff}}$ and the Laughlin state, remain constant with increasing system size (Fig. \ref{fig:fidelity_scaling}(b-c)). 
This observation strengthens the smooth transferability and suggests that for $H_{\text{eff}}$ considered here, reproducing local physics with high accuracy, does not require large prefactors in the linear depth scaling of the HVA.
As such, our protocol can be extended sensibly to near-term quantum simulations of strongly correlated topological systems at scale.

Lastly, the Hamiltonian in Eq.~\eqref{eqn:LaughlinExactGSHam} exhibits both particle number conservation $\hat{N} = \sum_j \hat{n}_j$ and center-of-mass coordinate conservation $\hat{K} = \sum_j j \hat{n}_j \ (\text{mod } N)$. The unitaries $\hat{U}_{km}$ composing our state preparation circuit naturally respect these symmetries, constraining the subspace of the variational search. Similarly, the final state $\ket{\psi(\{\beta_j\})}_{\text{eff}}$ must transform identically under these symmetries as the initial state $\ket{\Psi_0}$, enabling symmetry-verification protocols for robust error-mitigation \cite{mcardle_error-mitigated_2019, stanisic_observing_2022}.

\section{Edge and Bulk Density Structure}

\begin{figure}
    \centering
    \includegraphics[width=0.95\linewidth]{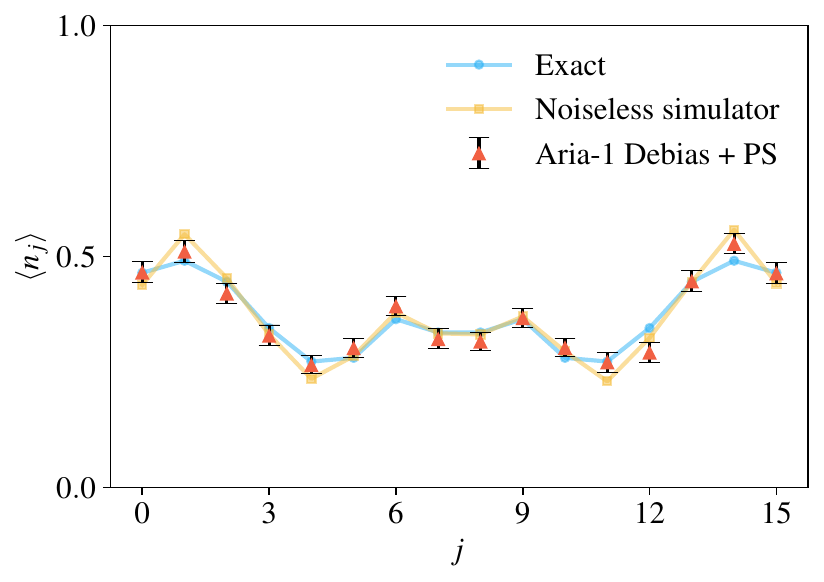}
    \caption{\textbf{Probing edge and bulk density structure.} $\langle n_j\rangle$ is the observed electron occupation at site $j$, obtained by sampling 5000 shots on IonQ's Aria-1 quantum computer with  symmetry-verification postselection (PS) and debiasing error-mitigation (red triangle), which leads to a $10\%$ selection rate. Error bars indicate $68\%$ confidence intervals obtained by means of percentile bootstrap. These results are compared with noiseless simulation of state preparation circuit (orange square) and exact values obtained by ED (blue circle). Lines are guide to the eye.}
    \label{fig:density_experiments}
\end{figure}

We next proceed to prepare and probe the Laughlin state on quantum processors. A key question we sought to address was whether a deep quantum circuit, involving hundreds of two-qubit gates but only a few variational parameters, could successfully capture the physics of strongly correlated topological states on NISQ devices. While the cost of storing and manipulating many-body wavefunctions grows exponentially on classical hardware, this experiment, if successful, would be an important step toward scalable quantum simulations for materials-intrinsic topological order on near-term quantum processors. Given the depth of the circuit, i.e., 369 two-qubit gates for $N_e = 6$, we selected a trapped-ion quantum processor (IonQ's 25-qubit Aria-1) for its relatively high two-qubit gate fidelity and low readout error rates, both of which are critical for mitigating noise and enabling effective post-selection strategies (see Methods).

One of the defining features of the quantum Hall states is the existence of chiral edge modes. On the cylinder geometry, the bulk-boundary correspondence \cite{wen1992theory, dubail2012edge} guarantees the presence of chiral edge modes, which emerge from the bulk’s nontrivial topological order and appear as oscillatory deviations in the local density structure near the physical boundary \cite{rezayi1994laughlin}. We can directly probe these edge structure in the prepared state by measuring the local density operator $\langle n_j\rangle = \langle c_j^\dagger c_j \rangle$ where $n_j = \frac{1}{2}(1 - Z_j)$ under Jordan-Wigner transformation.

In Fig. \ref{fig:density_experiments}, we present the measured $\langle n_j\rangle$ obtained by executing our state preparation circuit for $N_e = 6$ on Aria-1. Despite the limitation of current NISQ devices, the edge density structure is distinctly identified with an overdensity near the system boundaries ($j = 0, 15$) and subsequent oscillatory deviations of $\langle n_j\rangle$ from the bulk filling fraction $\nu = 1/3$. Away from the boundaries, the bulk region exhibits a relatively uniform density plateau, signaling the incompressibility and homogeneity nature of the topologically ordered Laughlin state. This spatial structure - a compressible, gapless edge surrounding an incompressible bulk - is an emblematic signature of FQH liquids.

The ability to resolve these edge structures relies critically on the symmetry-verification error mitigation that is naturally supported by our state preparation circuit. On the day of execution, Aria-1 reports a mean two-qubit gate fidelity of 98.5\%. With approximately 300 two-qubit gates per qubit's light-cone, a naive estimate implies a circuit fidelity of 1\%, making error mitigation crucial to retrieve meaningful information from experiments on NISQ device. To address this challenge, we employ a combined error mitigation strategy: a custom symmetry-verification postselection protocol alongside IonQ’s debiasing mitigation scheme \cite{maksymov_enhancing_2023}. The postselection depends on the conservation of particle number and center-of-mass coordinate that are both respected by our state preparation circuit. Any measured bitstrings violating either of these two symmetries are deemed unphysical and thus discarded during postselection.

With IonQ's debiasing mitigation alone, the result displays a systematic drift towards $\langle n_j\rangle = 0.5$, corresponding to the expectation value from a maximally mixed state, though the overall trend aligns qualitatively with the exact value obtained by ED. The application of symmetry-verification postselection significantly improves the fidelity of the results, eliminating the drift and confirming the observation of Laughlin state's edge density structure (see Supplementary Information for debiasing only data and details on postselection).

\begin{figure}
    \centering
    \includegraphics[width=0.95\linewidth]{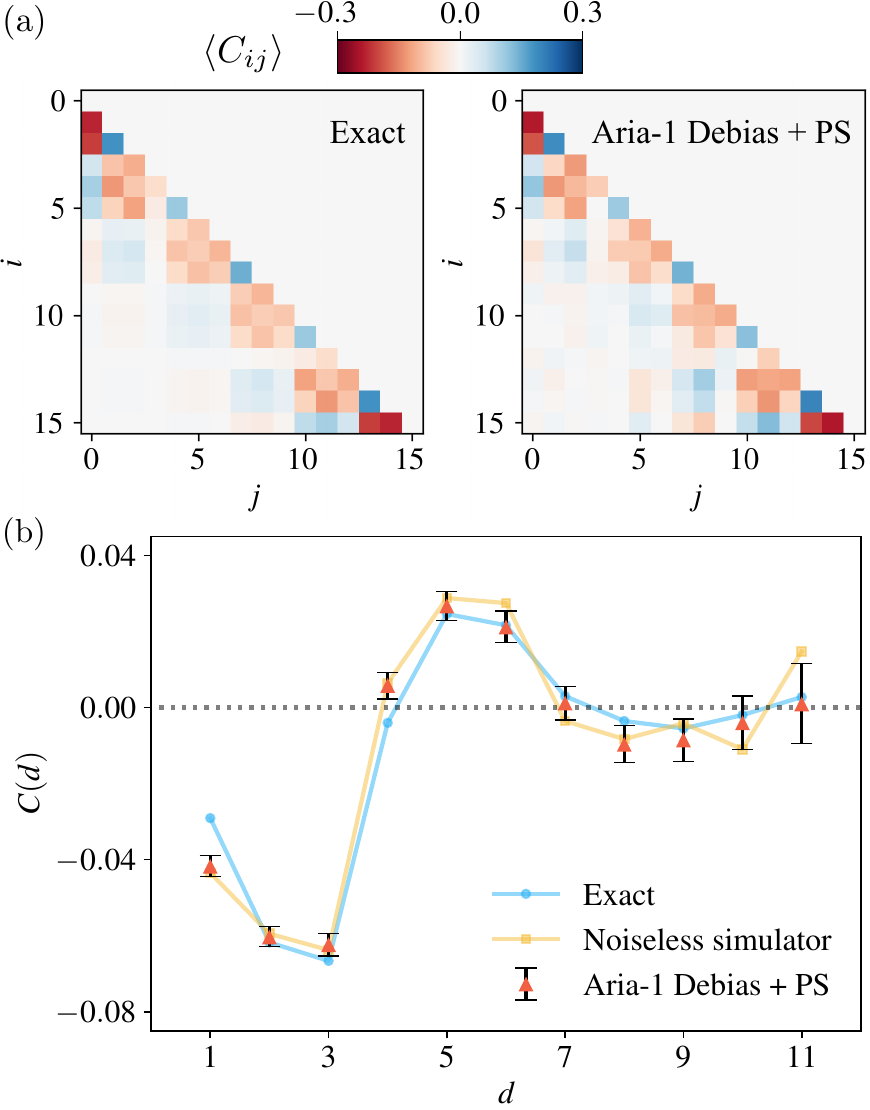}
    \caption{\textbf{Spatial correlations and incompressibility of the prepared Laughlin state. (a)} Two-point correlation function $C_{ij}$ between site $i$ and $j$ obtained from results after debiasing and postselection (PS) closely align with ED benchmark. We set $C_{ij} = 0$ for $i \le j$. \textbf{(b)} Site-averaged correlation $C(d)$ over sites separated by $d = |i - j|$. We include only site index $i,j \in [2, 13]$ when calculating $C(d)$ to avoid boundary effect. Error bars indicate $68\%$ confidence intervals obtained by means of percentile bootstrap. Lines are guide to the eye.}
    \label{fig:correlation_experiments}
\end{figure}

\section{Spatial Correlation and Topological Entanglement Entropy}

After establishing the presence of edge modes, we turn to investigate the incompressible bulk region of the prepared Laughlin state. In the bulk region, the Laughlin state behaves as an interacting incompressible quantum liquid. This results in a uniform featureless bulk density but leaves nontrivial spatial fingerprints in the wavefunction. To investigate such spatial characteristics, we measure the two-point correlation function $C_{ij} = \langle n_i n_j \rangle - \langle n_i \rangle \langle n_j \rangle$ between site $i$ and $j$. By construction, $C_{ij}$ is inversion-symmetric, that is, $C_{ij} = C_{ji}$ and approaches 1 ($-1$) when the electron densities are correlated (anticorrelated).

With debiasing mitigation alone, we observe clear spatial signatures of anticorrelation in the first two off-diagonal elements of $C_{ij}$, consistent with repulsive interactions (see Supplementary Information). After applying symmetry-verification postselection (Fig. \ref{fig:correlation_experiments}(a)), we fully resolve the spatial correlation contrast of the correlated electron liquid. Additionally, long-wavelength density fluctuations are strongly suppressed as $C_{ij}$ converges rapidly to zero as $|i-j|$ increases. The long-range correlation remains negligible in the bulk, except near the system's boundaries where edge effects dominate.

We further compute the site-averaged correlation function $C(d) = \overline{C_{j,j+d}}$ as a function of the separation distance $d = |i - j|$ and observe characteristic fluctuations in the short-range correlation of the prepared Laughlin state. The first two sites near each boundary are excluded to minimize edge effects. The results, shown in Fig. \ref{fig:correlation_experiments}(b), reveal a strong correlation hole $C(d) < 0$ at short distances ($d < 4$), signifying the underlying repulsive nature of Laughlin state. The medium-range oscillations in $C(d)$ reflect a short-range solid-like order, characteristic of a strongly coupled plasma. Such oscillations are a hallmark of the strongly correlated FQH liquid \cite{girvin_modern_2019}. Beyond $d \ge 7$, $C(d)$ decays rapidly to zero, representing a featureless and homogeneous liquid at long range. Not only does $C(d)$ from our prepared Laughlin state exhibit qualitative agreement across all distance ranges, but it also quantitatively captures the precise maxima and minima, as well as the spatial extent of the correlation hole. 

\begin{figure}
    \centering
    \includegraphics[width=0.95\linewidth]{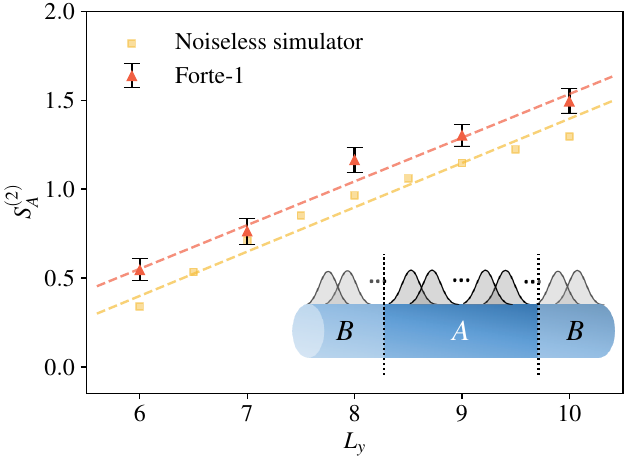}
    \caption{\textbf{Topological entanglement entropy.} The second-order R\'enyi entropy $S_A^{(2)}$ of the six-qubit subsystem as a function of cylinder circumference $L_y$. Red triangles represent experimental data obtained on IonQ's Forte-1 quantum computer using randomized measurements with an ensemble size of $N_U=200$ unitaries and $N_M=300$ shots per unitary. The result is compared with noiseless simulation of the variationally optimized HVA (orange square). Dashed lines indicate linear fits to the area law form $S_2(L_y) = \alpha L_y - \gamma$. The noiseless simulation yields $\alpha_\text{HVA}=0.249$ and $-\gamma_{\text{HVA}}=-1.09$. Experimental fit yields $\alpha_\text{exp}= 0.245 \pm 0.021$ and $-\gamma_\text{exp}=-0.92 \pm 0.17$. Error bars indicate 68\% confidence intervals obtained by means of percentile bootstrap. \textbf{Inset:} Schematic of the orbital partition. The system is partitioned into a bulk subsystem $A$ and the environment $B$, illustrating the two spatial cuts contributing to the entanglement entropy.}
    \label{fig:renyi_entropy}
\end{figure}

To demonstrate entanglement behavior beyond pairwise correlation, we directly measured the topological entanglement entropy $\gamma_\text{topo}$ \cite{kitaev2006topological, levin_detecting_2006} of our prepared state via geometric deformation of the cylinder circumference $L_y$ on the quantum processor.
This quantity, which reflects the quantum dimension of anyonic excitations, serves as a robust diagnostic of topological order.
We optimized the HVA ansatz $\ket{\psi(\{\beta_j\})}_{\text{eff}}$ for a range of $L_y \in [6, 10]$ near the isotropic geometry limit, and applied a randomized measurement protocol \cite{brydges_probing_2019} to estimate the second-order R\'enyi entropy $S^{(2)}_A = - \ln \text{Tr}\,\rho_A^2$ for three different subsystem partition $A$ in the bulk region. (see Supplementary Information for details)

In Fig. \ref{fig:renyi_entropy}, the experimentally measured $S^{(2)}_A$ shows the expected area-law scaling $S^{(2)}_A = \alpha L_y - \gamma_\text{topo}$ with a systematic drift to higher entropy due to hardware noise when compared to noiseless simulator benchmark.
Fitting the measured second-order R\'enyi entropy to the area-law scaling, we extracted $-\gamma_\text{exp} = -0.92 \pm 0.17$ (68\% confidence interval by bootstrap resampling of finite-shot randomized measurement estimator, see Methods).
For the ideal $\nu = 1/3$ Laughlin state, $-\gamma_\text{topo} = -\ln \sqrt{3}$ \cite{zozulya2007bipartite, haque2007entanglement, lauchli2010entanglement} and because our system partition introduces two entanglement boundaries, the expected value is $-\gamma_\text{topo} = -2 \ln \sqrt{3} \approx -1.10$.
The consistent behavior in $S^{(2)}_A$ and $\gamma_\text{topo}$ between our experiments and the theory provides compelling evidence of the topological order of the prepared $\nu = 1/3$ Laughlin state. 
Our pairwise correlation and entanglement entropy measurements demonstrate the ability to access microscopic structures that underlies topologically ordered states on a quantum processor.

\section{Discussion and Outlook}

In summary, we realized a strongly correlated topological order on IonQ’s trapped-ion quantum computer by preparing the $\nu = 1/3$ Laughlin state using an efficient and scalable HVA. 
We validated our experiment by extracting FQH phase-diagnostic observables from hardware and sets a benchmark for future experiments on larger, classically intractable systems.

Beyond the Laughlin state, our method can be extended to quasiparticle states \cite{seidel_domain-wall-type_2007} as well as to more complex non-Abelian topological order such as the Moore–Read \cite{moore_nonabelions_1991} and Read–Rezayi \cite{read_beyond_1999} states. The realization of these exotic phases would mark a significant step towards exploring exotic topological phases, providing a robust platform for exploring Abelian and non-Abelian braiding statistics through adiabatic quasiparticle transport \cite{flavin_abelian_2011}, edge and bulk excitations, and nonequilibrium dynamics such as emergent graviton modes in FQH systems \cite{haldane2011geometrical}. Moreover, the ability to prepare these exotic states position our approach as a promising testbed for benchmarking next-generation quantum processors.

In addition, our work demonstrates a hardware-validated end-to-end workflow for studying strongly correlated topological materials. Unlike classical methods, which are fundamentally constrained by exponential complexity, our quantum simulation workflow provides a scalable route to access key properties of these systems such as phase stability, low-energy excitations, and response functions by directly preparing and probing these states on quantum processors.
Moreover, our protocol is well positioned as a practical state-initialization routine for a broader class of quantum algorithms \cite{berry_rapid_2025, yoshioka_krylov_2025, huggins_unbiasing_2022, he_alignment_2023} where high-quality initial states substantially improve algorithm's convergence and practical performance. Taken together, these features position our protocol as a versatile building block for the digital simulation and diagnosis of topological quantum matter.

\section{Acknowledgments}

We thank Pavel Lougovski, Xiaodong Hu, and Xiaodong Xu for insightful discussions. The authors acknowledge the support of Amazon Web Services (AWS) credit and quantum computing resources for this work. The computations and simulations of this work are supported by the U.S. Department of Energy, Office of Basic Energy Sciences, under Contract No. DE-SC0025327. The theoretical analysis is supported by the U.S. Department of Energy, Office of Science, National Quantum Information Science Research Centers, Co-design Center for Quantum Advantage (C2QA), under Contract No. DE-SC0012704. This research used resources of the National Energy Research Scientific Computing Center, a DOE Office of Science User Facility supported by the Office of Science of the U.S. Department of Energy under Contract No. DE-AC02-05CH11231 using NERSC award BES-ERCAP0032546, BES-ERCAP0033507, and DDR-ERCAP0034430.

\section{Methods}

\subsection{Fractional quantum hall Hamiltonian}
We consider two-dimensional (2D) interacting electron gas subject to a perpendicular magnetic field $B$ on a cylinder geometry, where $L_x$ and $L_y$ denote the length and circumference respectively, and $N_\Phi = L_x L_y / (2\pi)$ specifies the total number of magnetic flux quanta threading the cylinder. For finite cylinder geometries, the number of flux quanta satisfies $N_\Phi = 3N_e - 2$, where $N_e$ is the number of electrons \cite{misguich_dmrg_2021, rezayi1994laughlin}. Throughout this work, we set the magnetic length $l_B \equiv \sqrt{\frac{\hbar}{eB}}$ to unity for simplicity. Under the Landau gauge $\vec{A} = Bx \hat{y}$ where $\hat{y}$ is the direction of the circumference of the cylinder, the problem is reduced from a 2D continuum system to an effective one-dimensional (1D) lattice model. For spinless electrons within the $l$-th Landau level, the two-body interaction assumes the following 1D lattice model \cite{seidel_incompressible_2005, moudgalya_quantum_2020}
\begin{equation}
    H_l = \sum_{j_1, j_2, j_3,j_4} V_{j_1, j_2, j_3,j_4}^{(l)}c^\dagger_{j_1}c^\dagger_{j_2} c_{j_3} c_{j_4},
\end{equation}
where $c^\dagger_j$ and $c_j$ are the fermionic creation and annihilation operators for single-particle orbital $\psi_{l,j}(\mathbf{r})$ with $j$ being the index for both the $\hat{x}$ center-of-mass coordinate and the $\hat{y}$ momentum eigenvalue. For example, the associated single-particle orbital for the lowest Landau level ($l = 0$) on a cylinder is
\begin{equation}
    \psi_{0,j}(\mathbf{r}) = \frac{1}{\sqrt{L_y \sqrt{\pi}}} e^{i y\frac{2 \pi}{L_y}j} e^{-(x - \frac{2 \pi}{L_y}j)^2 /2}.
\end{equation}

The matrix element $V_{j_1, j_2, j_3,j_4}^{(l)}$ is obtained by projecting the two-body interaction onto the space spanned by $\psi_{l,j}(\mathbf{r})$.
The Hamiltonian $H_l$ can be further simplified to
\begin{equation}
    H_{l} = \sum_j \sum_{k>m} V_{km}^{(l)} c^\dagger_{j+m}c^\dagger_{j + k} c_{j + k + m} c_{j}.
\end{equation}

To study the $\nu=1/3$ Laughlin state, we focus on the lowest Landau level and adopt the Haldane-Trugman-Kivelson pseudopotential \cite{haldane_fractional_1983, trugman_exact_1985}
\begin{equation}
    V(\mathbf{r}_1 - \mathbf{r}_2) = \nabla^2\delta(\mathbf{r}_1 - \mathbf{r}_2),
\end{equation}
which guarantees the $\nu=1/3$ Laughlin state as an exact ground state. The corresponding matrix elements in LLL are given by \cite{moudgalya_quantum_2020}
\begin{equation}
    V_{km}^{(0)} = \frac{16 \pi^2}{L_y} (k^2 - m^2) e^{-\frac{2 \pi^2 (k^2+m^2)}{L_y^2}}.
\end{equation}
which physically represents a short-ranged repulsion in the guiding center coordinates that penalizes electrons being too close.

\subsection{Efficient Hamiltonian variational ansatz}

Hybrid quantum-classical algorithms \cite{peruzzo_variational_2014, mcclean_theory_2016, cerezo_variational_2021, stanisic_observing_2022, hemery_measuring_2024, nigmatullin_experimental_2024} provide a viable strategy for quantum simulations in the NISQ era by employing shallow, parameterized circuits refined through classical optimization. 
Among the proposed approaches, HVA has emerged as a promising candidate \cite{wecker_progress_2015}. Consider a general Hamiltonian,
\begin{equation}
H = \sum_j c_j \hat{h}_j,
\end{equation}
where $c_j$ are scalars and $\hat{h}_j$ are operators. A single repetition of HVA is constructed using unitary evolution operators,
\begin{equation}
    \ket{\psi(\{\beta_j\})} = \prod_j \exp(-i \beta_j \hat{h}_j) \ket{\Psi_0},
\end{equation}
where $\beta_j$ are variational parameters and $\ket{\Psi_0}$ is an initial state that can be easily prepared. The variational parameters are classically optimized against a loss function. This flexibility allows state preparation with much shallower circuit compared to circuit mimicking a trotterized annealing processes.

After decomposing the correlated topological electronic Hamiltonian
\begin{equation}
    H = H_{\text{eff}} + H',
\end{equation}
where $H_{\text{eff}}$ is an effective Hamiltonian retaining the essential interactions and $H'$ contains the subdominant contributions.
The corresponding Hamiltonian variational ansatz \cite{wecker_progress_2015} constructed from $H_\text{eff}$ is
\begin{equation}
    \ket{\psi(\{\beta_j\})}_{\text{eff}} = \prod_j \exp(-i \beta_j \hat{h}_j) \ket{\Psi_0},\ \hat{h}_j \in H_\text{eff}.
\end{equation}
This approach reduces computational complexity while preserving both quantitative accuracy and qualitative topological features. Unlike models that target topologically trivial phases, in which Hubbard-like on-site interaction terms are usually sufficient to describe electron-electron interactions, our method retains long-range interactions crucial for nontrivial topological order, improving both expressiveness and physical fidelity.

An additional advantage of this approach lies in its preservation of Hamiltonian symmetries. By construction, the symmetry constraints ensure that the final state $\ket{\psi(\{\beta_j\})}_\text{eff}$ transforms under the same symmetry operations by $H$ as the initial state, regardless of variational parameters $\{\beta_j\}$. This property enables the ansatz to target ground states associated with specific quantum numbers, determined by the initial state $\ket{\Psi_0}$. In addition, such symmetry requirement confines the optimization to the physically relevant subspace, reducing classical search complexity while enabling symmetry-verification error mitigation on quantum hardware \cite{mcardle_error-mitigated_2019}.

\subsection{Variational optimization procedure}

The HVA associated with $H_{\text{eff}}$ is given by
\begin{equation}
    \label{eqn:ansatz_laughlin}
    \ket{\psi(\{\beta_j\})}_{\text{eff}} = \hat{U}_{20}\hat{U}_{10}\hat{U}_{31}\hat{U}_{30}\hat{U}_{21} \ket{\Psi_{\text{CDW}}},
\end{equation}
where the CDW state $\ket{\Psi_{\text{CDW}}} = \ket{100100....1001}$ serves as the initial state, prepared by applying $X$ gates on every three qubits of the trivial product state $\ket{0}^{\otimes N}$.

The variational optimization problem is formulated as 
\begin{equation}
 \min_{\{\beta_{km}\}}  \langle H \rangle(\{\beta_{km}\}) = \bra{\psi(\{\beta_j\})}_\text{eff} H \ket{\psi(\{\beta_j\})}_\text{eff},
\end{equation}
where $H$ denotes the parent Hamiltonian for Laughlin state Eq.~\eqref{eqn:LaughlinExactGSHam}. Optimization of ${\beta_{km}}$ was performed via classical simulation (noiseless). Specifically, for a fixed system size $N$, we optimize the expectation value of $H$ at filling factor $\nu = 1/3$ in the isotropic geometry regime, setting the circumference $L_y = 10$, where the system's ground state is the Laughlin state. 
We used the \texttt{pennylane.lightning} package to perform a noiseless simulation of the ansatz circuit and output the exact quantum state vector and the \texttt{numpy} package to compute the expectation value of $H$.

We use the L-BFGS-B algorithm for optimization, as implemented in the \texttt{SciPy} package \cite{byrd1995limited, zhu1997algorithm}, combined with basinhopping to mitigate the risk of converging to local minima. The basinhopping routine was performed with $10^2$ hopping attempts, and each local optimization was allowed a maximum of $10^3$ iterations. To further enhance robustness, we initialized the optimization from 50 independent random initial parameter sets. Convergence was declared when the relative change in the cost function, $\langle H \rangle$, was less than $10^{-6}$ between successive iterations.

\begin{table}
\centering
\begin{tabular}{c c c c c}
 \hline
 $\beta_{21}$ & $\beta_{30}$ & $\beta_{31}$ & $\beta_{10}$ & $\beta_{20}$\\ [0.5ex]
 \hline\hline
 11.751 & 12.573 & 12.219 & 4.732 & 10.972 \\ [1ex]
 \hline
\end{tabular}
\caption{Optimized parameters for $\nu = 1/3$ Laughlin state at $L_y = 10$ with system size $N_e = 6$.}
\label{table:optimized_beta}
\end{table}

\subsection{Gate decomposition for scattering layer $\hat{U}_{km} (m \ne 0)$}

Implementing the scattering layer $\hat{U}_{km} (m \ne 0)$ on a quantum processor requires efficient decomposition into native gate operations. In this work we adopt the first-order Suzuki-Trotter method to implement all the unitaries $\hat{U}_{km}$. After Jordan-Wigner transformation, the exponent in
\begin{equation}
    \hat{U}_{km} = \prod_{j} \text{exp}[-i \beta_{km} (c^\dagger_{j+m}c^\dagger_{j + k} c_{j + k + m} c_{j} + \text{H.c.})],
\end{equation}
for a specific $j$ will yield 8 Pauli terms
\begin{align}
    XYXY, YY&XX, XXXX, YXXY, \nonumber \\
    &XYYX, YYYY, XXYY, YXYX
\end{align}
where we have omitted the qubit index for conciseness. Reordering these terms strategically can significantly reduce the circuit depth by minimizing basis changes between successive Trotter steps. We rearrange them as follows
\begin{align}
    XXXX, XX&YY, XYXY, XYYX, \nonumber \\
    &YYXX, YYYY, YXXY, YXYX
\end{align}
This optimized sequencing leads to a substantial constant factor reduction in CNOT gate overhead, decreasing the count from 48 to 17 per site index $j$. We used \texttt{qiskit} for circuit compilation.

\subsection{Statistical inference and error propagation}

For each circumference $L_y$, the second R\'enyi entropy estimator $S_A^{(2)}(L_y)$ was obtained from randomized measurements using an ensemble of $N_U=200$ random unitaries and $N_M=300$ shots per unitary. To reduce partition-dependent finite-size oscillations and improve statistical efficiency, we evaluated three different subsystem partitions of size $N_A=6$ and pooled their primitive purity estimators prior to the logarithm. Denoting by $X_{m,u}(L_y)$ the primitive estimator for cut $m\in\{1,2,3\}$ and unitary $u$, we form
\begin{equation}
Y_u(L_y)=\frac{1}{3}\sum_{m=1}^3 X_{m,u}(L_y),\qquad
\widehat{P}(L_y)=\frac{1}{N_U}\sum_{u=1}^{N_U} Y_u(L_y).
\end{equation}
We report the bias-corrected entropy estimator
\begin{equation}
\widehat{S}_A^{(2)}(L_y)= -\ln \widehat{P}(L_y)\;-\;\frac{\widehat{\mathrm{Var}}(\widehat{P}(L_y))}{2\,\widehat{P}(L_y)^2},
\end{equation}
where the second term is a second-order delta-method correction for the nonlinear $-\ln(\cdot)$ transformation and
$\widehat{\mathrm{Var}}(\widehat{P}(L_y))=s_Y^2(L_y)/N_U$ with
$s_Y^2(L_y)=\frac{1}{N_U-1}\sum_u \left(Y_u(L_y)-\widehat{P}(L_y)\right)^2$.

Pointwise uncertainty at fixed $L_y$ was quantified by a non-parametric bootstrap over the unitary ensemble ($10^4$ replicates). In each replicate we resampled the $N_U$ unitaries with replacement and recomputed $\widehat{S}_A^{(2),*}(L_y)$ using the same pooled-purity and bias-correction procedure (with the same resampled unitary indices applied to all three partitions to preserve their correlations). We plot the means of the resulting bootstrap distribution, with error bars defined by the 16th and 84th percentiles.

To propagate uncertainty to the linear-fit parameters in $S_A^{(2)}(L_y)=\alpha L_y-\gamma_{\mathrm{topo}}$, we performed a refit bootstrap ($10^4$ replicates). For each replicate, we generated $\{\widehat{S}_A^{(2),*}(L_y)\}_{L_y}$ as above and fit by weighted least squares with fixed weights $w_i=1/\sigma_i$, where $\sigma_i$ is the standard deviation of the pointwise bootstrap distribution at $L_{y,i}$. The reported $\alpha$ and $-\gamma_{\mathrm{topo}}$ are the means of the resulting parameter distributions, with uncertainties obtained from the 16th and 84th percentiles (symmetrized for compact reporting).

\subsection{Quantum hardware}

The quantum circuits for measuring local density and spatial correlations were executed on IonQ’s Aria 1 trapped-ion quantum computer, which utilizes 25 ytterbium-ion-based qubits with all-to-all connectivity. The hardware is calibrated daily and here we report Aria 1’s calibrations on the day of execution accessed through Amazon Braket. Single-qubit gates were characterized using Clifford randomized benchmarking, achieving an average fidelity of 99.97\%. Two-qubit gates were benchmarked using direct randomized benchmarking on the $XX(\pi/4)$ gate, yielding an average fidelity of 98.46\%. Readout fidelity was evaluated through one-qubit randomized benchmarking, with an average fidelity of 99.44\%. To mitigate hardware noise, we employed IonQ’s native debiasing mitigation scheme.

The quantum circuits for measuring second-order R\'enyi entropy via randomized measurement were executed on IonQ’s Forte 1 trapped-ion quantum computer, which utilizes 36 ytterbium-ion–based qubits with all-to-all connectivity. Here we report Forte 1’s characterization on the day of execution accessed through Amazon Braket. Single-qubit gates were characterized using Clifford randomized benchmarking, achieving an average fidelity of 99.98\%. Two-qubit gates were benchmarked using direct randomized benchmarking \cite{chen_benchmarking_2024}, yielding an average fidelity of 99.68\%. State preparation and measurement (SPAM) was evaluated through one-qubit randomized benchmarking, with an average fidelity of 99.68\%.

\bibliography{citations}

\end{document}


\title{Supplementary Information: Realization of fermionic Laughlin state on a quantum processor}

\author{Lingnan Shen}
    \affiliation{Department of Physics, University of Washington, Seattle, WA, USA}

\author{Mao Lin}
    \affiliation{Amazon Braket, Seattle, WA, USA}
\author{Cedric Yen-Yu Lin}
    \affiliation{Amazon Braket, Seattle, WA, USA}

\author{Di Xiao}
    \email{dixiao@uw.edu}
    \affiliation{Department of Material Science and Engineering, University of Washington, Seattle, WA, USA}
    \affiliation{Department of Physics, University of Washington, Seattle, WA, USA}
    \affiliation{Pacific Northwest National Laboratory, Richland, WA, USA}
\author{Ting Cao}
    \email{tingcao@uw.edu}
    \affiliation{Department of Material Science and Engineering, University of Washington, Seattle, WA, USA}
\date{\today}

\maketitle
\tableofcontents

\section{Hamiltonian under the Tao-Thouless limit}

The system Hamiltonian is given by
\begin{equation}
    \label{eqn:2quantizedParentHam}
    H = \sum_j \sum_{k>m} V_{km} c^\dagger_{j+m}c^\dagger_{j + k} c_{j + k + m} c_{j},
\end{equation}
with the Haldane-Trugman-Kivelson psuedopotential matrix element
\begin{equation}
    V_{km} = \frac{16 \pi^2}{L_y} (k^2 - m^2) e^{-\frac{2 \pi^2 (k^2+m^2)}{L_y^2}}.
\end{equation}
In this formulation, the scattering terms characterized by nonzero values of $m$ ($m \ne 0$) become exponentially suppressed compared to the dominant electrostatic interaction terms ($V_{k0}$) in the Tao-Thouless limit $L_y \to 0$. Thus the system becomes a charge-density-wave (CDW) state $\ket{\Psi_{\text{CDW}}} = \ket{100100100...}$.

The lowest order approximation beyond the Tao-Thouless (TT) limit is made by truncating long-range interactions up to $(k + m) \le 3$ terms, yielding the TT limit Hamiltonian 
\begin{equation}
    \label{eqn:SI_H_TT}
    H_{\text{TT}} = \sum_j [V_{10} \hat{n}_j\hat{n}_{j+1} + V_{20} \hat{n}_j\hat{n}_{j+2}  + V_{30} \hat{n}_j\hat{n}_{j+3} + (V_{21} c^\dagger_{j+1}c^\dagger_{j + 2} c_{j + 3} c_{j} + \text{H.c.})],
\end{equation}
which includes only the lowest-order scattering term $V_{21} c^\dagger_{j+1}c^\dagger_{j + 2} c_{j + 3} c_{j}$.

\subsection{Krylov subspace formation}

The fidelity decay at isotropic limit for $H_{\text{TT}}$ with $(k+m \leq 3)$ originates from Hilbert space fragmentation, where the system is confined to a Krylov subspace $\mathcal{K}$ defined as
\begin{equation}
    \mathcal{K} \equiv \text{Span} \{ \ket{\Psi_{0}}, H_{\text{TT}} \ket{\Psi_{0}}, H_{\text{TT}}^2 \ket{\Psi_{0}}, ...\},
\end{equation}
the subspace connected to the root charge-density-wave state $\ket{\Psi_0} = \ket{\Psi_{\text{CDW}}} = \ket{100100100...}$ by the action of $H_{\text{TT}}$. This Krylov subspace $\mathcal{K}$ is significantly \textit{smaller} than the full Hilbert space of Laughlin state
\begin{equation}
    \mathcal{H}_{\text{TT}} = \mathcal{K} \subset \mathcal{H}_{\text{Laughlin}},\ \text{dim}(\mathcal{K}) \ll \text{dim}(\mathcal{H}_{\text{Laughlin}}).
\end{equation}

Thus, $H_{\text{TT}}$ remains valid only near the TT limit and fails to capture the relevant correlations in the full Laughlin state and its underlying topology. Including higher-order scatterings like $V_{31}$ in $H_\text{eff}$ breaks the constraint of $\mathcal{K}$, connecting the entire Hilbert space with $\ket{\Psi_0}$ and ensuring the expressiveness of the Hamiltonian variational ansatz.

\subsection{Entanglement entropy}

We further analyze the validity of different truncation range by comparing the R\'enyi entanglement entropy $S_A$ of their ground states. The second-order R\'enyi entanglement entropy is defined as
\begin{equation}
    S_A^{(2)} = -\ln \text{Tr}\rho_A^2,
\end{equation}
where $\rho_A = \text{Tr}_B(\ket{\Psi} \bra{\Psi})$ is the reduced density matrix for subsystem A by tracing over the degrees of freedom of subsystem B. In the following calculation, we choose $A$ to be the left half orbitals of the cylinder.

For a system in $d$ dimensions with a finite correlation length $l$, the entanglement entropy satisfies the area law
\begin{equation}
    S_A \simeq \alpha L^{d-1},
\end{equation}
where $L$ is the length of the boundary between the two blocks. For the two-dimensional Laughlin state, we expect $S_A$ to scale linearly with the cylinder's circumference $L_y$.

\begin{figure}[h]
    \centering
    \includegraphics[width=0.45\linewidth]{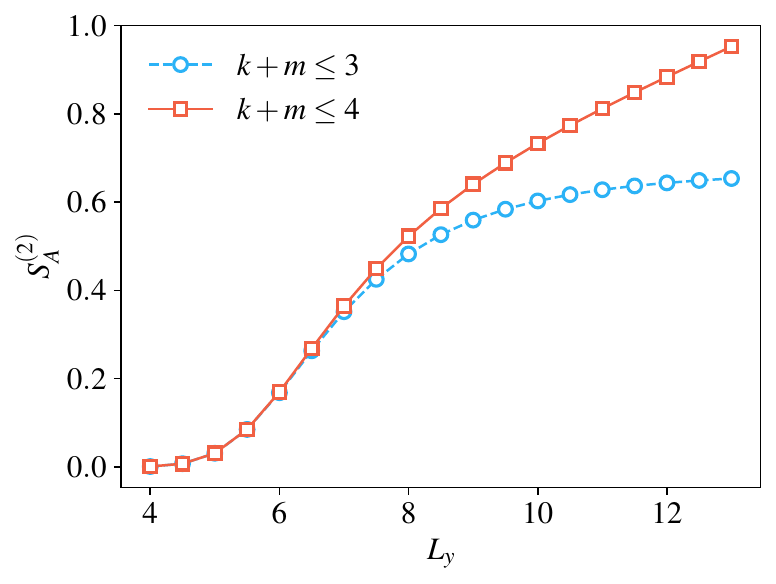}
    \caption{Second R\'enyi entanglement entropy $S_A^{(2)}$ of the ground state of the effective Hamiltonian for various truncation ranges of interactions ($k+m\le 3, 4$) on a finite cylinder geometry, as a function of cylinder circumference $L_y$. The system size is $N_e = 8$, and the bipartition is chosen to divide the system into two equal halves, each containing 11 orbitals.}
    \label{fig:SI_vn_entropy}
\end{figure}

As shown in Fig. \ref{fig:SI_vn_entropy}, in the TT limit ($L_y \to 0$), the system ground state approaches the charge-density-wave state $\ket{\Psi_{\text{CDW}}} = \ket{100100100...}$ which is a product state, leading to $S_A \to 0$. In the region $L_y \lesssim 7$, the entanglement entropy of the ground state for both interaction truncation $(k+m \le 3, 4)$ align closely. However, beyond $L_y > 7$, the entropy of the ground state of $H_\text{TT}$ quickly saturates, deviating from the expected area-law scaling of a genuine fractional quantum Hall liquid. This indicates that $H_{\text{TT}}$ fails to capture relevant entanglement in the full Laughlin state and its underlying topology. In contrast, extending the truncation range to $(k+m \le 4)$ recovers the expected area law behavior for an incompressible topological quantum liquid.

\section{Additional results of Hamiltonian variational ansatz optimization}

\subsection{Optimizing Hamiltonian variational ansatz correspond to $H_{\text{TT}}$}

In addition, we optimized the Hamiltonian variational ansatz correspond to the Tao-Thouless limit Hamiltonian $H_{\text{TT}}$ in Eq. \ref{eqn:SI_H_TT} for system size $N_e = 6$
\begin{equation}
    \ket{\psi(\{\beta_j\})}_{\text{TT}} = \hat{U}_{20}\hat{U}_{10}\hat{U}_{30}\hat{U}_{21} \ket{\Psi_{\text{CDW}}}.
\end{equation}
The highest fidelity achieved was $\mathcal{F}=0.79$, significantly lower than the $\mathcal{F}=0.93$ obtained for $\ket{\psi({\beta_j})}_{\text{eff}}$ in the main text. 

\subsection{Optimizing $\ket{\psi(\{\beta_j\})}_{\text{eff}}$ at $N_e = 8$}

We also optimized the Hamiltonian variational ansatz $\ket{\psi({\beta_j})}_{\text{eff}}$, as defined in the main text, for larger system size $N_e = 8$. The highest fidelity achieved through direct optimization was $\mathcal{F}=0.88$. Remarkably, this optimized fidelity closely matches the fidelity obtained by extrapolating parameters previously optimized at $N_e = 6$ to larger system size $N_e = 8$.

\subsection{Optimization result for ansatz with $V_{40}$ term}
We included an additional term $\hat{U}_{40}$ and performed optimization at system size $N_e = 6$. Specifically, we considered the variational ansatz:
\begin{equation}
    \ket{\psi(\{\beta_j\})}_{\text{TT}} = \hat{U}_{40}\hat{U}_{20}\hat{U}_{10}\hat{U}_{30}\hat{U}_{21} \ket{\Psi_{\text{CDW}}}.
\end{equation}
Upon optimization, the maximum fidelity achieved was $\mathcal{F} = 0.929$, only marginally improved compared to the ansatz without the $\hat{U}_{40}$ term ($\mathcal{F} = 0.927$). Thus the $H_\text{eff}$ defined in main text represent the minimum effective Hamiltonian for constructing the HVA for quantum simulation of $\nu = 1/3$ Laughlin state. 

\section{Detailed resource required by Hamiltonian variational ansatz circuit}

\subsection{Example Hamiltonian variational ansatz circuit for $N_e = 6$}

We present an example quantum circuit for a 16-qubit system, corresponding to $N_e = 6$, which was executed on IonQ’s Aria-1 trapped-ion quantum processor.
\begin{figure}[h]
    \centering
    \includegraphics[width=0.9\linewidth]{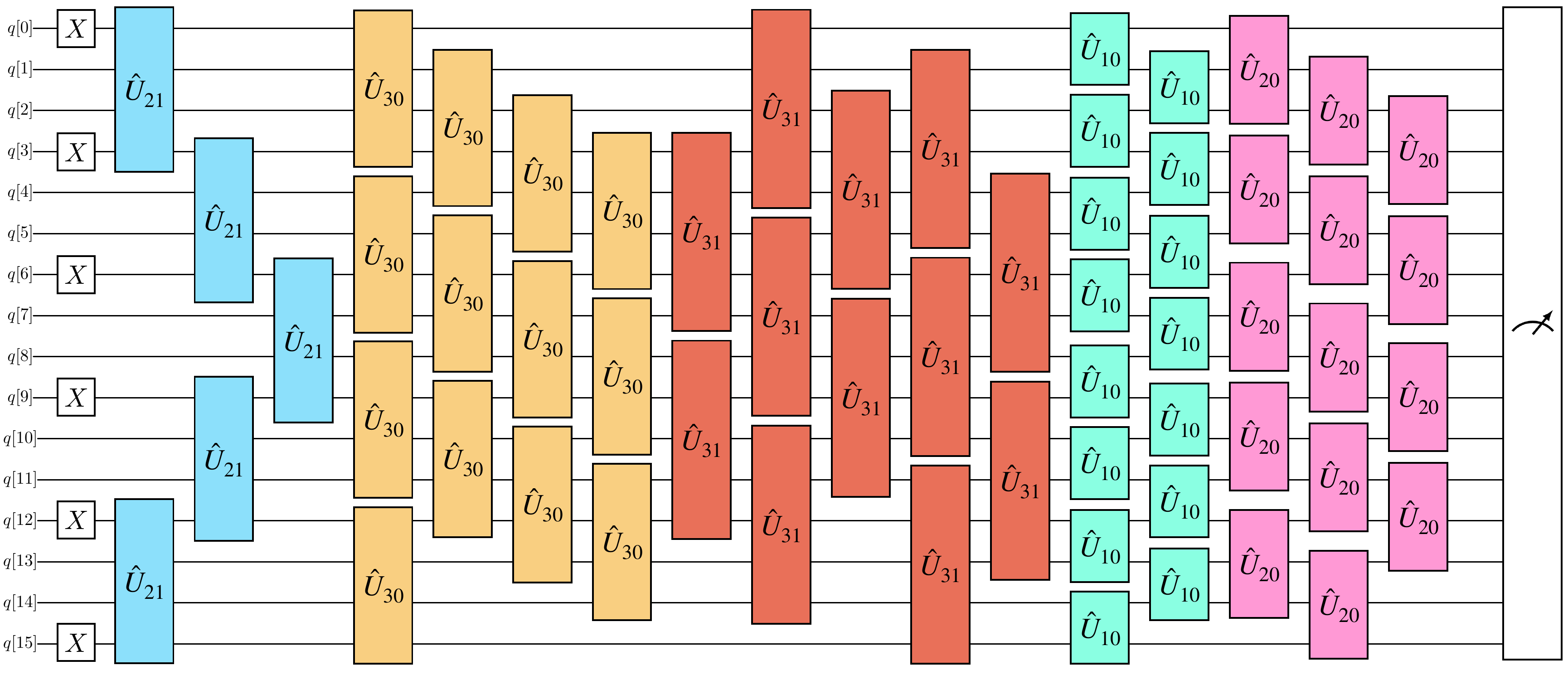}
    \caption{Example circuit for the 16-qubit HVA corresponding to the $\nu = 1/3$ Laughlin state for system size $N_e = 6$.}
    \label{fig:SI_circuit_Ne_6}
\end{figure}

\subsection{Circuit gate count and depth scaling}

Here, we analyze how the circuit two-qubit (CNOT) gate count and depth scale with both the interaction truncation of the Hamiltonian and the system size. The Hamiltonian variational ansatz constructed with respect to Eq.~\ref{eqn:2quantizedParentHam} is
\begin{equation}
    |\psi \rangle = \prod_{k>m} \hat{U}_{km} |\Psi_{\text{CDW}} \rangle, \quad \hat{U}_{km} = \prod_{j} \exp\big[-i \beta_{km} (c^\dagger_{j+m} c^\dagger_{j + k} c_{j + k + m} c_{j} + \text{H.c.})\big],
\end{equation}
where the interaction truncation range is defined by $k + m \le \text{cutoff}$. The CNOT gate count (Table \ref{table:gate_counts}) grows roughly linearly with system size and exponentially with the cutoff while the CNOT depth (Table \ref{table:gate_depth}) scales approximately exponentially with the truncation cutoff and only weakly with system size. We show in Table \ref{table:HVA_resource} the quantum resources needed for a single repetition of the HVA we constructed in main text.

\begin{table}[h]
    \centering
    \begin{tabular}{ccccc}
        \toprule
        Cutoff &\ $N_e = 6$ &\ $N_e = 8$ &\ $N_e = 10$ &\ $N_e = 12$ \\
        \hline\hline
         3 &  305   &  443   &  581   &   719   \\
         4 &  533   &  785   & 1037   &  1289   \\
         5 & 1156   & 1752   & 2335   &  2929   \\
         6 & 1722   & 2660   & 3585   &  4516   \\
         7 & 2775   & 4411   & 6035   &  7675   \\
         8 & 3714   & 6048   & 8375   & 10713   \\
         9 & 5053   & 8537   & 11999  & 15482   \\
        10 & 6081   & 10593  & 15076  & 19591   \\
        \hline
    \end{tabular}
    \caption{CNOT gate count with respect to interaction range cutoff and system size $N_e$.}
    \label{table:gate_counts}
\end{table}

\begin{table}[h]
    \centering
    \begin{tabular}{ccccc}
        \toprule
        Cutoff &\ $N_e = 6$ &\ $N_e = 8$ &\ $N_e = 10$ &\ $N_e = 12$ \\
        \hline\hline
         3 &  86   &  86   &  86   &  86   \\
         4 & 177   & 179   & 181   & 181   \\
         5 & 507   & 515   & 515   & 520   \\
         6 & 892   & 906   & 906   & 912   \\
         7 & 1805  & 1820  & 1821  & 1832  \\
         8 & 2722  & 2848  & 2851  & 2860  \\
         9 & 4031  & 4722  & 4722  & 4729  \\
        10 & 5039  & 6564  & 6560  & 6576  \\
        \hline
    \end{tabular}
    \caption{CNOT depth with respect to interaction range cutoff and system size $N_e$.}
    \label{table:gate_depth}
\end{table}

\begin{table}[h]
    \centering
    \begin{tabular}{>{\centering\arraybackslash}m{3em} >{\centering\arraybackslash}m{5em} >{\centering\arraybackslash}m{6em}} 
        \toprule
        $N_e$ & Qubits & CNOT gates \\ [0.5ex]
        \hline\hline
        6 & 16 & 369 \\
        8 & 22 & 543 \\
        10 & 28 & 711 \\
        12 & 34 & 883 \\ [0.5ex] 
        \hline
    \end{tabular}
    \caption{Number of qubits and CNOT gates for a single repetition of our HVA $\ket{\psi(\{\beta_j\})}_{\text{eff}}$ for various system size $N_e$. The number scales approximately linearly with the system size.}
    \label{table:HVA_resource}
\end{table}

\section{Measurement of topological entanglement entropy}

\subsection{Extracting topological entanglement entropy $\gamma_\text{topo}$ through geometric deformation}

To isolate the topological contribution to the entanglement, we analyze the scaling behavior of the second-order R\'enyi entropy $S_A^{(2)}$ of the variationally optimized ground state $|\psi(\{\beta_j\})\rangle_{\text{eff}}$ using the relation
\begin{equation}
    S_A^{(2)} \simeq \alpha L_y - \gamma_\text{topo},
\end{equation}
where $\alpha$ is a non-universal constant and $L_y$ is the circumference of the cylinder. 
Since the $\nu=1/3$ Laughlin phase is gapped, its ground state correlations decay exponentially and the entanglement associated with an orbital cut is localized within a finite correlation length around each entanglement boundary. 
Previous analysis on the cylinder give a bulk correlation length $\zeta_{1/3}\approx 1.38\,\ell_B$ for the Laughlin state \cite{crepel2019microscopic}, which in the Landau-gauge orbital basis corresponds to an orbital correlation length $\xi_{\rm orb}\equiv \zeta/\Delta x \sim (\zeta/\ell_B)(L_y/2\pi\ell_B)$ (with $\Delta x=2\pi\ell_B^2/L_y$), i.e. $\xi_{\rm orb} \leq 3$ orbitals over our experimental near-isotropic regime of $L_y$.
Accordingly, we choose the subsystem $A$ (containing $N_A = 6$ orbitals) around the bulk region of the cylinder, see Sec.~\ref{sec:entropy_SI_extended_detail}, that is larger than $\xi_{\rm orb}$ while keeping the randomized-measurement sampling cost manageable (as the number of measurements required to estimate purities grows exponentially with subsystem size \cite{brydges_probing_2019}).
This introduces two disjoint entanglement boundaries between subsystem $A$ and the environment $B$. Consequently, the topological correction is doubled, leading to a theoretical prediction of: $-\gamma_\text{topo} = -2\ln\sqrt{3} \approx 1.10$.

To benchmark our variational ansatz, we optimized $|\psi(\{\beta_j\})\rangle_{\text{eff}}$ for $N_e = 6$ electrons on a classical simulator across a range of cylinder circumferences $L_y \in [5.5, 10.0]$. We computed $S_A^{(2)}$ for each optimized state and performed a linear fit to the area law scaling form. The extracted intercept from the classical simulation is $-\gamma_{\text{vqe}} = -1.09$ by bootstrap refitting over three system partitions' entropy (see Sec.~\ref{sec:entropy_SI_extended_detail} for detail), which stands in excellent agreement with the theoretical prediction, validating the capability of the ansatz to capture the essential topological correlations of the Laughlin phase.

To verify that the entanglement entropy obtained from $|\psi(\{\beta_j\})\rangle_{\text{eff}}$ arises from intrinsic topological order rather than simple density modulation, we variationally optimized a minimal excitation-preserving ansatz constrained to match only the local density $\langle n_j \rangle$ profile as obtained from $|\psi(\{\beta_j\})\rangle_{\text{eff}}$ at each $L_y$. Despite achieving a density profile of $ \max_j \Delta \langle n_j \rangle < 0.05$ from $|\psi(\{\beta_j\})\rangle_{\text{eff}}$, the minimal excitation-preserving ansatz exhibits a significantly smaller $S^{(2)}_A$.

\subsection{Randomized measurement protocol for second-order Rényi entropy}

To experimentally access the topological entanglement entropy, we measured the second-order Rényi entropy $S_A^{(2)} = -\ln \text{Tr}(\rho_A^2)$ of the same six-qubit subsystem $A$. We employed the randomized measurement protocol \cite{brydges_probing_2019}, which estimates the purity $X_A = \text{Tr}(\rho_A^2)$ from the statistical correlations of bitstring outcomes in randomized local bases, avoiding the exponential overhead of full quantum state tomography.

For each cylinder circumference $L_y$ we prepared the optimized HVA and applied an ensemble of $N_U = 200$ independent random unitaries. Each unitary $U^{(r)} = \bigotimes_{i=1}^{N} U_i^{(r)}$ consisted of local single-qubit gates drawn from the Circular Unitary Ensemble. Following the application of $U^{(r)}$, the system was measured in the computational basis, yielding a set of $N_M = 300$ bitstrings per unitary. For a subsystem $A$ of size $N_A$, the estimated purity $X_A^{(r)}$ by a random unitary $U^{(r)}$ is given by:
\begin{equation}
    X_A^{(r)} = 2^{N_A} \sum_{s_A,s_A'} (-2)^{-D(s_A, s_A')} P(s_A) P(s_A'),
\end{equation}
where $s_A$ and $s_A'$ denote bitstring outcomes restricted to the subsystem $A$, $D(s_A, s_A')$ is the Hamming distance between them and $P(s_A)$ denotes the probability of observing $s_A$. In practice, $P(s_A)^2$ is a biased estimator for $\mathbb{E}(P(s_A))^2$ due to finite shot noise. Instead, we calculated the purity using an unbiased estimator
\begin{equation}
    X_A^{(r)} = 2^{N_A} \left[\sum_{s_A \ne s_A'} (-2)^{-D(s_A, s_A')} \frac{N_M}{N_M-1}P(s_A) P(s_A') + \sum_{s_A}  \frac{P(s_A)(N_M P(s_A) -1)}{N_M - 1}\right].
\end{equation}
The estimated purity is then acquired by averaging over the unitary ensemble 
\begin{equation}
    \text{Tr} (\rho_A^2) = \frac{1}{N_U} \sum_{r=1}^{N_U} X_A^{(r)}.
\end{equation}
In this work, we performed non-parametric bootstrap analysis on the ensemble of $N_U$ purity estimates to determine second R\'enyi entropy.

To mitigate the effects of hardware drift and fluctuations during execution, we implemented an interleaved execution scheme. Rather than collecting all $N_M$ shots for a given unitary at once, we cycled through the full unitary ensemble in small batches to accumulate the total measurement shots, improving effective stability without changing $N_U$ and $N_M$.

\section{Quantum hardware data}

\subsection{IonQ Aria-1 data}

\begin{figure}
    \centering
    \includegraphics[width=0.45\linewidth]{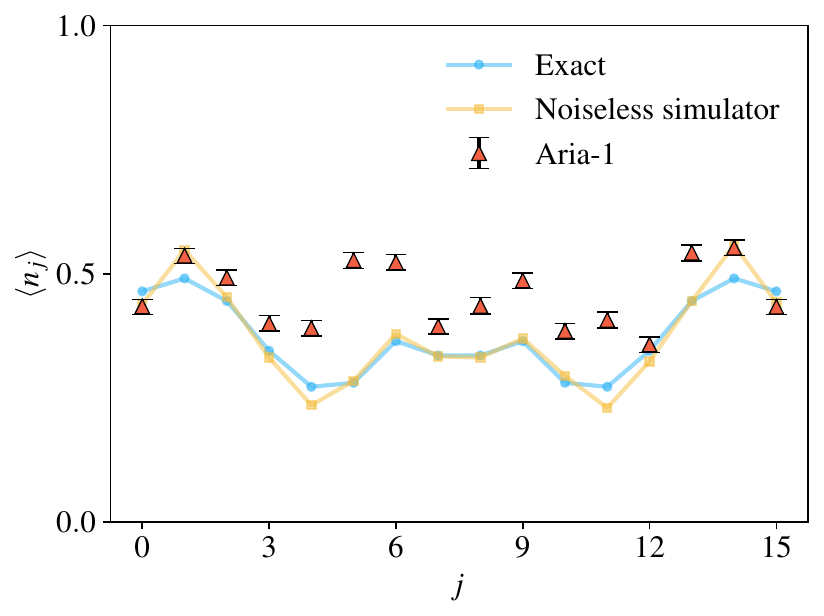}
    \caption{Local density $n_j$ at site $j$ for system size $N_e = 6$. Experimental results obtained without error-mitigation effort are compared with noiseless simulation of the optimized HVA (orange square) and exact values from ED (blue circle). Error bars indicate $68\%$ confidence intervals obtained by means of percentile bootstrap.}
    \label{fig:SI_n_j_no_mitigation}
\end{figure}

\begin{figure}
    \centering
    \includegraphics[width=0.45\linewidth]{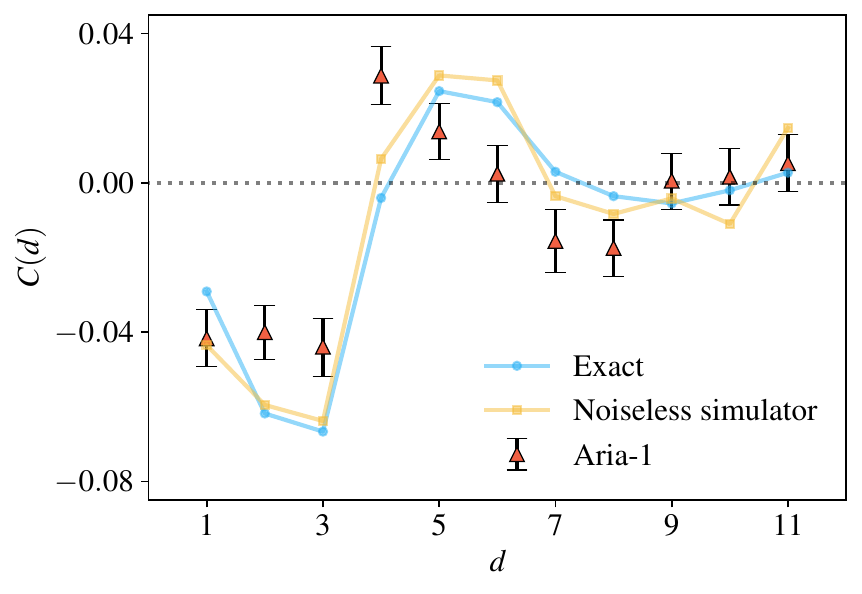}
    \caption{Site-averaged correlation $C(d)$ over sites separated by $d = |i - j|$. Experimental results obtained without error-mitigation effort are compared with noiseless simulation of the optimized HVA (orange square) and exact values from ED (blue circle). Error bars indicate $68\%$ confidence intervals obtained by means of percentile bootstrap.}
    \label{fig:SI_c_ij_d_no_mitigation}
\end{figure}

We sampled 1000 shots on IonQ's Aria-1 quantum computer. Experimental results without using any error-mitigation technique are shown for local density $n_j$ (Fig. \ref{fig:SI_n_j_no_mitigation}) and site-averaged correlation $C(d)$ (Fig. \ref{fig:SI_c_ij_d_no_mitigation}).

\subsection{IonQ Aria-1 debiasing data}
\begin{figure}[h]
    \centering
    \includegraphics[width=0.45\linewidth]{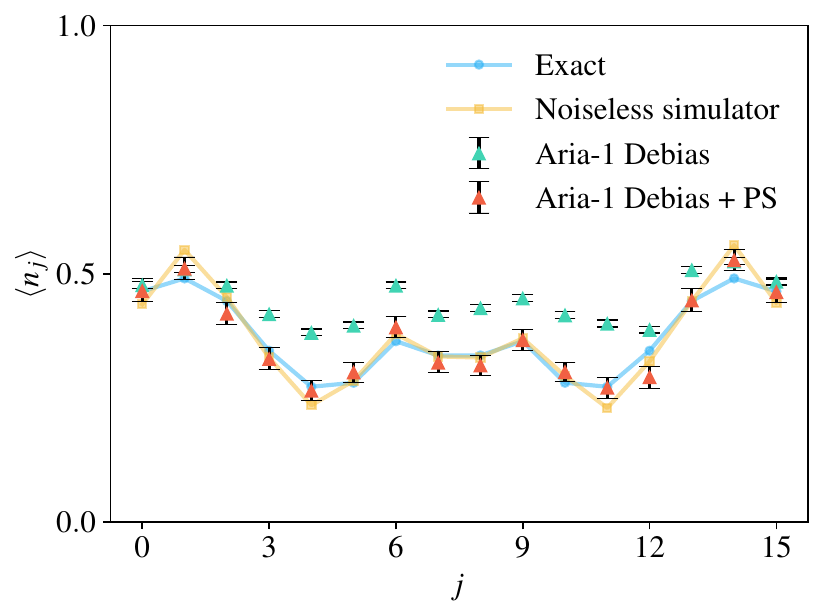}
    \caption{Local density $n_j$ at site $j$ for system size $N_e = 6$. Experimental results with debiasing error mitigation (cyan triangle) and symmetry-verification postselection (red triangle) are compared with noiseless simulation of the optimized HVA (orange square) and exact values from ED (blue circle). Error bars indicate $68\%$ confidence intervals obtained by means of percentile bootstrap.}
    \label{fig:SI_n_j}
\end{figure}

\begin{figure}[h]
    \centering
    \includegraphics[width=0.35\linewidth]{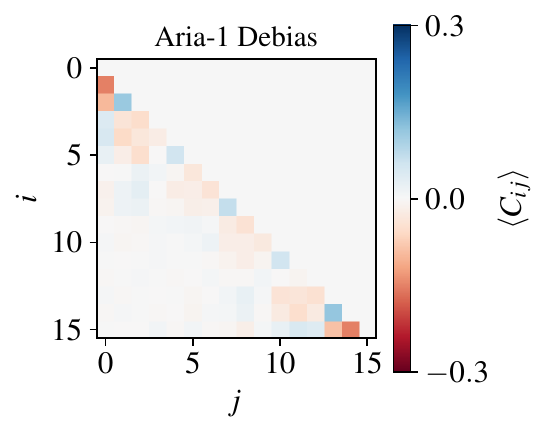}
    \caption{Two-point correlation function $C_{ij}$ between site $i$ and $j$ with debiasing error-mitigation alone. We set $C_{ij} = 0$ for $i \le j$ for clarity.}
    \label{fig:SI_c_ij}
\end{figure}

\begin{figure}[h]
    \centering
    \includegraphics[width=0.45\linewidth]{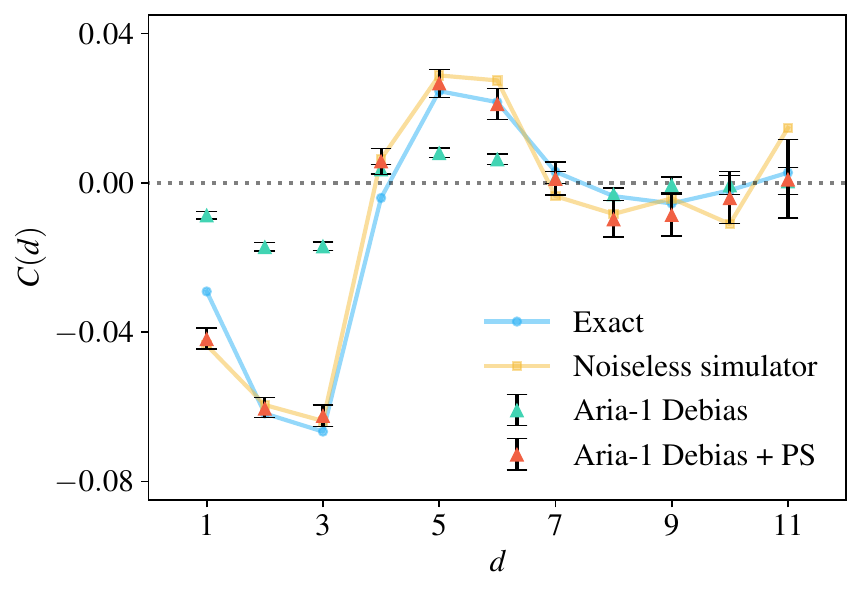}
    \caption{Site-averaged correlation $C(d)$ over sites separated by $d = |i - j|$. Experimental results with debiasing error mitigation (cyan triangle) and symmetry-verification postselection (red triangle) are compared with noiseless simulation of the optimized HVA (orange square) and exact values from ED (blue circle). Error bars indicate $68\%$ confidence intervals obtained by means of percentile bootstrap.}
    \label{fig:SI_c_ij_d}
\end{figure}

We sampled 5000 shots on IonQ's Aria-1 quantum computer. Experimental results using IonQ’s debiasing mitigation alone is shown for local density $n_j$ (Fig. \ref{fig:SI_n_j}), two-point correlation function $C_{ij}$ (Fig. \ref{fig:SI_c_ij}), and site-averaged correlation $C(d)$ (Fig. \ref{fig:SI_c_ij_d}). While all experimental data follows the general qualitative trend of the ED benchmark, we still see large deviations due to hardware noise.

\subsection{Symmetry-verification post-selection}

To further mitigate errors arising from quantum hardware execution, we employ a symmetry-verification postselection scheme. This method discards measurement bitstrings that violate the conservation of particle number, $\hat{N} = \sum_j \hat{n}_j$, and center-of-mass position, $\hat{K} = \sum_j j \hat{n}_j \ (\text{mod } N_{\Phi})$. Specifically, only measurement bitstrings satisfying the following conditions are retained
\begin{equation}
    N = 6,\ K= 13.
\end{equation}

\begin{figure}[h]
    \centering
    \includegraphics[width=0.4\linewidth]{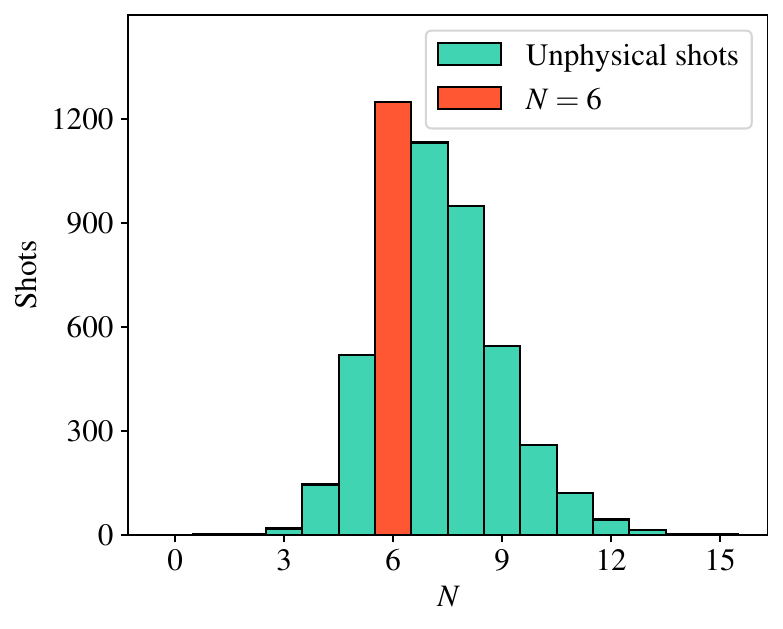}
    \caption{Distribution of measurement bitstrings by particle number $N$. 24.9\% of total shots (red) satisfy the particle number conservation ($N=6$), while the remaining measurement bitstrings (cyan) are deemed unphysical and discarded.}
    \label{fig:SI_particleNum_dist}
\end{figure}

\begin{figure}[h]
    \centering
    \includegraphics[width=0.4\linewidth]{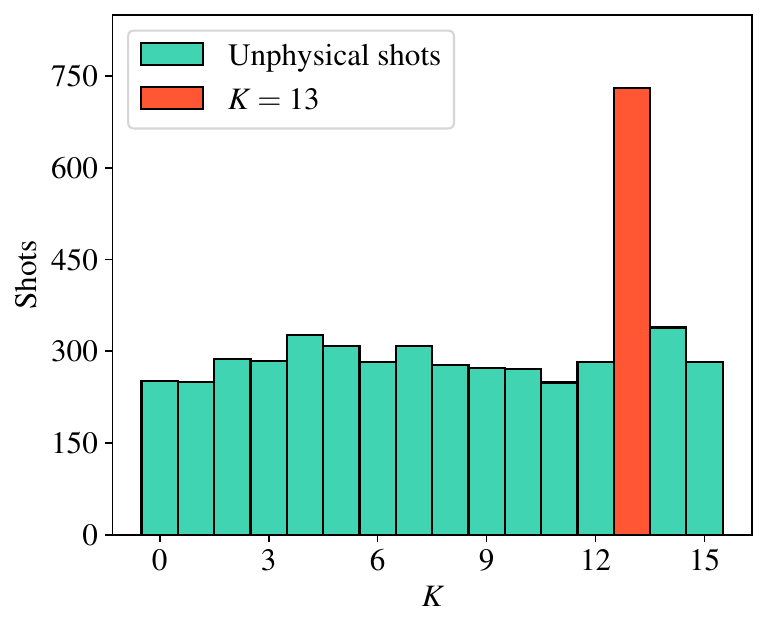}
    \caption{Distribution of measurement bitstrings by center-of-mass position. Only 14.6\% of total shots (red) satisfy the center-of-mass conservation ($K=13$), while the remaining measurement bitstrings (cyan) are deemed unphysical and discarded.}
    \label{fig:SI_com_dist}
\end{figure}

Fig. \ref{fig:SI_particleNum_dist} and Fig. \ref{fig:SI_com_dist} show the measurement bitstring distributions for particle number and center-of-mass position, respectively. After debiasing, 24.9\% of the measurements satisfy particle number conservation, while 14.6\% satisfy center-of-mass conservation. Enforcing both symmetries yields a final selection rate of 10.4\%.

\subsection{IonQ Forte-1 extended data of second Rényi entropy measurement}
\label{sec:entropy_SI_extended_detail}

\begin{figure}
    \centering
    \includegraphics[width=0.45\linewidth]{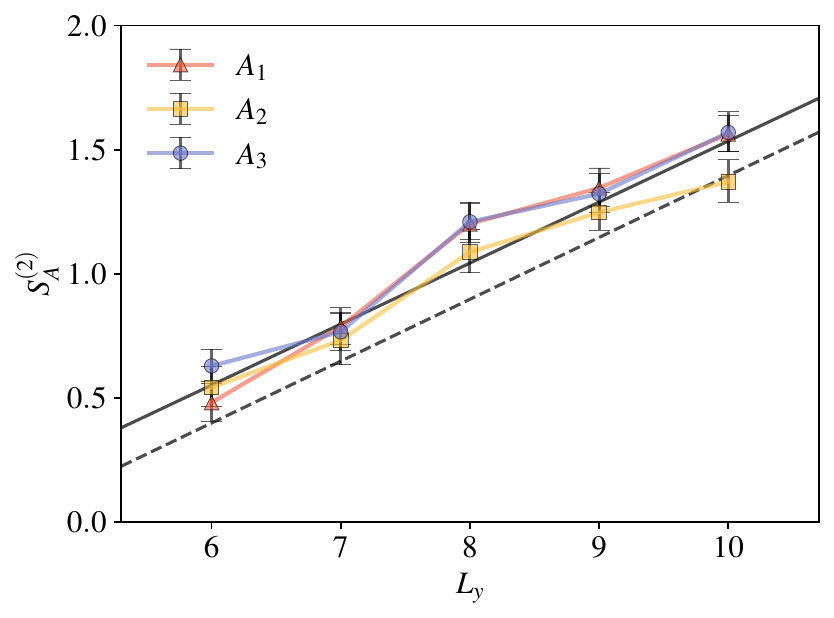}
    \caption{Second R\'enyi entropies for subsystem partitions $A_1,A_2,A_3$ measured on Forte-1. Error bars denote 68\% confidence intervals obtained by bootstrap. Colored lines are guide to eye. Area law fitted against Forte-1 measurements (black solid line) with $\alpha_\text{exp}= 0.245 \pm 0.021$ and $-\gamma_\text{exp}=-0.92 \pm 0.17$ and area law fitted against noiseless simulation (black dash line) with $\alpha_\text{HVA}=0.249$ and $-\gamma_{\text{HVA}}=-1.09$ are shown for reference.}
    \label{fig:SI_renyi_entropy_3_partitions}
\end{figure}

To suppress edge effects from the open boundaries during the extraction of second R\'enyi entropy $S_A^{(2)}$, we choose three contiguous subsystem blocks of fixed size $N_A=6$ centered around the middle of the cylinder, in the thin-torus root-pattern picture for $\nu=1/3$,
\begin{align}
    & A_1=1001|001001|001001,\\
    & A_2=10010|010010|01001,\\
    & A_3=100100|100100|1001,
\end{align}
where vertical bars indicate the subsystem blocks.
These three partitions are separated from both open ends by at least four orbitals and lie around the density plateau region observed in the main text, ensuring that the extracted entanglement is dominated by bulk physics rather than boundary reconstruction.
The resulting $S_A^{(2)}(L_y)$ data for each partition $A_1,A_2,A_3$ are shown in Fig.~\ref{fig:SI_renyi_entropy_3_partitions}.

For finite cylinders in the Landau-gauge orbital representation, short-distance physics at the subsystem partition boundaries can produce a weak position dependence in the nonuniversal part of $S_A^{(2)}$.
Previous work \cite{lauchli2010entanglement} explicitly demonstrates that, at fixed subsystem size, translated subsystem partitions can yield different entropy values, precisely because the microscopic boundary configuration controls the entanglement spectrum near the cut. They further employ an arithmetic average over translated partitions to suppress this oscillatory cut dependence and obtain a smoother scaling behavior.

To suppress this cut-dependent oscillations and improve statistical efficiency, we adopt the same arithmetic average over the three types of subsystem partition $A_1,A_2,A_3$.
In practice, we compute $S_A^{(2)}(L_y)$ for each of the three subsystem partitions and report the entropy obtained with means of bootstrap.

\begin{figure}
    \centering
    \includegraphics[width=1.0\linewidth]{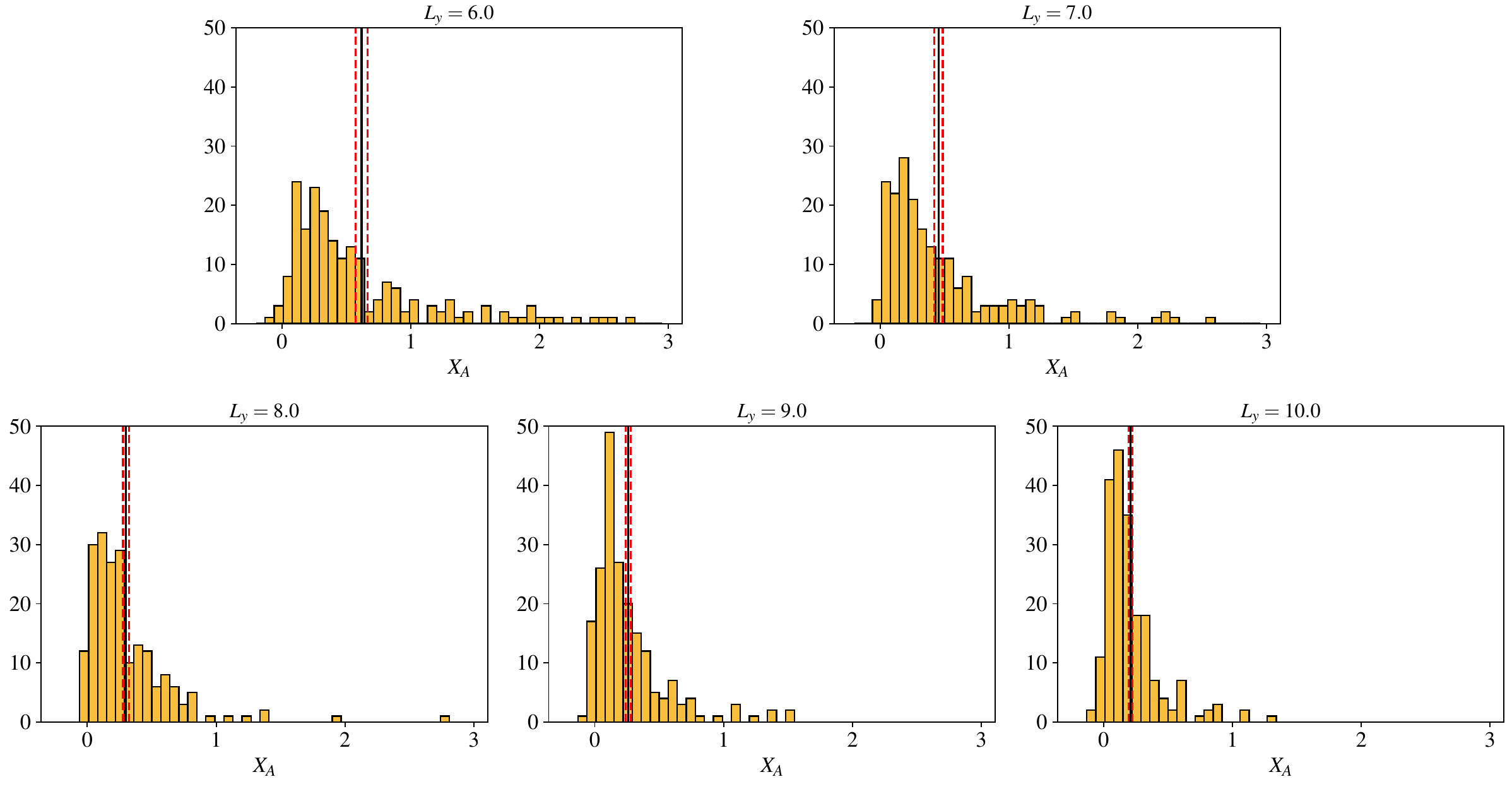}
    \caption{Distribution of purity estimators from randomized measurements. Histograms display the distribution of the purity estimator $X_A$ for the six-qubit bulk subsystem $A_1$ across the ensemble of $N_U = 200$ random unitaries for cylinder circumferences $L_y \in \{6, 7, 8, 9, 10\}$. Black line indicates the mean purity value obtained via bootstrapping, while the red dashed lines denote the 68\% confidence interval derived from percentile bootstrapping.}
\end{figure}

\begin{figure}
    \centering
    \includegraphics[width=1.0\linewidth]{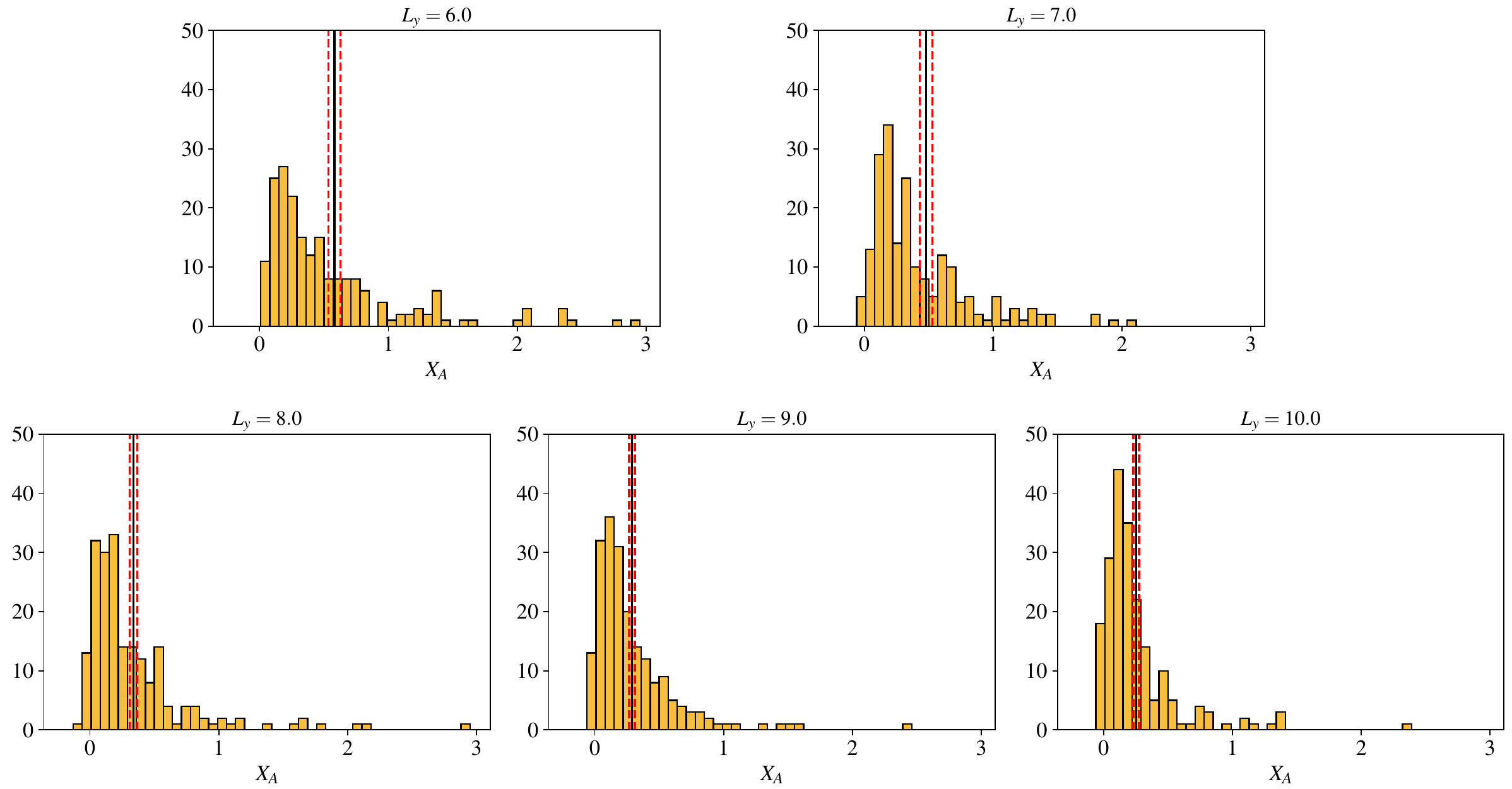}
    \caption{Distribution of purity estimators from randomized measurements. Histograms display the distribution of the purity estimator $X_A$ for the six-qubit bulk subsystem $A_2$ across the ensemble of $N_U = 200$ random unitaries for cylinder circumferences $L_y \in \{6, 7, 8, 9, 10\}$. Black line indicates the mean purity value obtained via bootstrapping, while the red dashed lines denote the 68\% confidence interval derived from percentile bootstrapping.}
\end{figure}

\begin{figure}
    \centering
    \includegraphics[width=1.0\linewidth]{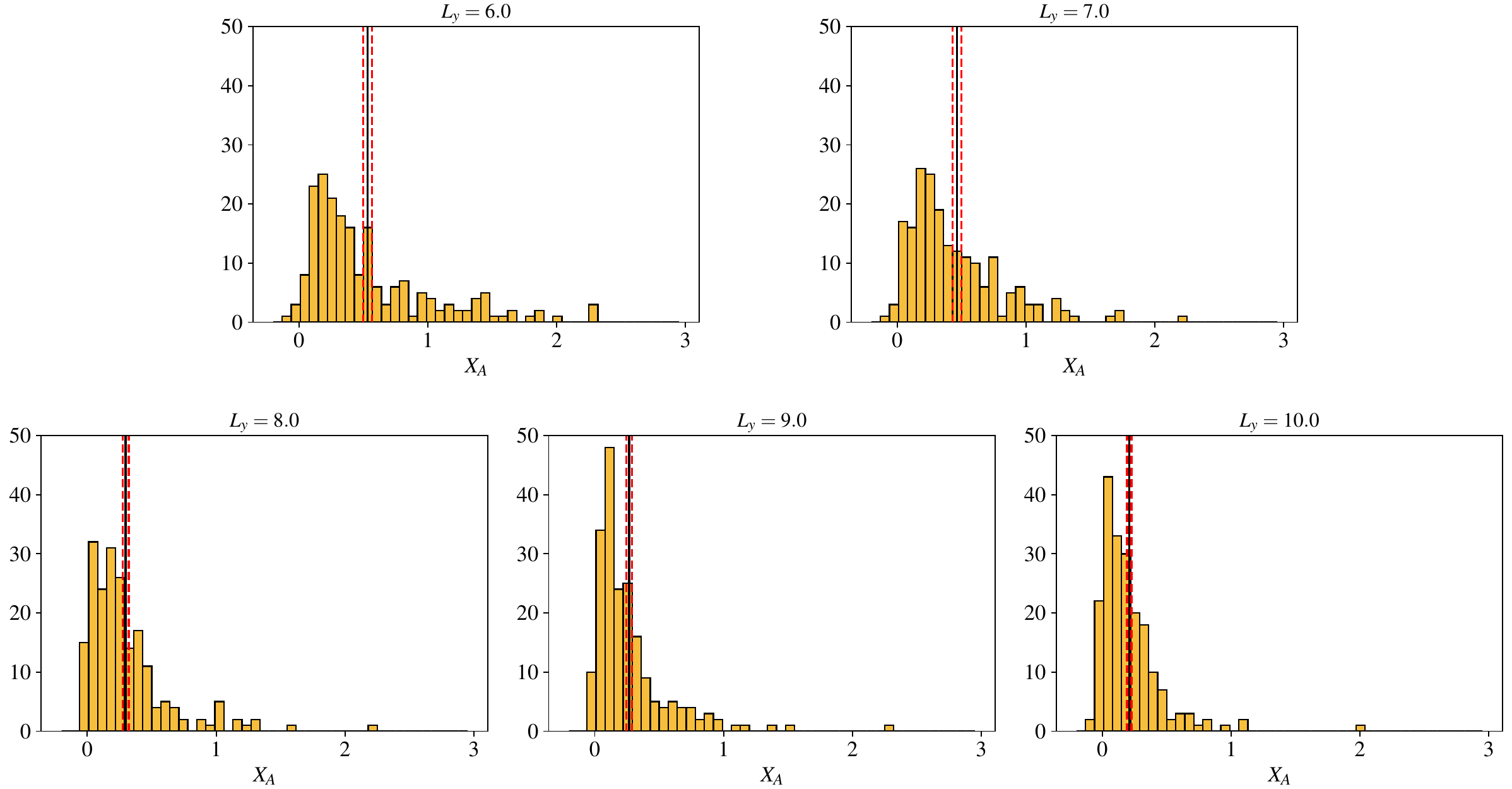}
    \caption{Distribution of purity estimators from randomized measurements. Histograms display the distribution of the purity estimator $X_A$ for the six-qubit bulk subsystem $A_3$ across the ensemble of $N_U = 200$ random unitaries for cylinder circumferences $L_y \in \{6, 7, 8, 9, 10\}$. Black line indicates the mean purity value obtained via bootstrapping, while the red dashed lines denote the 68\% confidence interval derived from percentile bootstrapping.}
\end{figure}

\bibliography{citations}